\documentclass[12pt]{article} \input epsf.tex
\usepackage{epsfig,graphicx,psfrag}
\usepackage{axodraw}
\usepackage{amsmath}\usepackage{amsfonts}
\usepackage{latexsym}\usepackage{stmaryrd}
\usepackage{amssymb}

\renewcommand{\theequation}{\thesection.\arabic{equation}}
\newcommand{\be}{\begin{equation}}
\newcommand{\ee}{\end{equation}}
\newcommand{\bea}{\begin{eqnarray}}
\newcommand{\eea}{\end{eqnarray}}
\newcommand{\bear}{\begin{eqnarray}}
\newcommand{\eear}{\end{eqnarray}}
\newcommand{\ba}{\begin{array}}
\newcommand{\ea}{\end{array}}
\newcommand{\lae}{\begin{array}{c}\,\sim\vspace{-21pt}\\<
\end{array}}
\newcommand{\gae}{\begin{array}{c}\,\sim\vspace{-21pt}\\>
\end{array}}

\newcommand{\beq}{\begin{equation}}
\newcommand{\eeq}{\end{equation}}
\newcommand{\beqs}{\begin{eqnarray}}
\newcommand{\eeqs}{\end{eqnarray}}

\newcommand{\myl}{\ensuremath{\lambda}}
\newcommand{\lp}{\ensuremath{\lambda^\prime}}

\newcommand{\he}{\ensuremath{\Phi_8}}
\newcommand{\hep}{\ensuremath{\Phi_8^\prime}}

\newcommand{\Mev}{ {\rm MeV} }
\newcommand{\gev}{ {\rm GeV} }
\newcommand{\tev}{ {\rm TeV} }

\newcommand{\inl}{{\scriptscriptstyle L}}
\newcommand{\inr}{{\scriptscriptstyle R}}
\newcommand{\inh}{{\scriptscriptstyle H}}

\newcommand{\dilog}{\ensuremath{\mathrm{Li}_2}}

\begin{document}

\baselineskip=18pt \pagestyle{plain} \setcounter{page}{1}

\vspace*{-0.8cm}

\noindent \makebox[11.6cm][l]{\small \hspace*{-.2cm} May 6, 2008; \, revised June 30, 2008 }{\small FERMILAB-Pub-08-49-T}  
\\ [-2mm]

\begin{center}
{\Large \bf Quark and lepton masses from top loops} \\ [9mm]

{\normalsize \bf Bogdan A. Dobrescu and Patrick J. Fox \\ [4mm]
{\small {\it
Theoretical Physics Department, Fermilab, Batavia, IL 60510, USA }}\\
}
\end{center}

\vspace*{0.1cm}

\begin{abstract}
Assuming that the leptons and quarks other than top are massless at tree level, 
we show that their masses may be induced by loops involving the top quark.
As a result, the generic features of the fermion mass spectrum arise
from combinations of loop factors. Explicitly, we construct a  
renormalizable model involving a few new particles, which leads to 
1-loop bottom and tau masses, a 2-loop charm mass,
3-loop muon and strange masses, and 4-loop masses for first generation fermions.
This realistic pattern of masses does not require any symmetry to 
differentiate the three generations of fermions.
The new particles may produce observable effects in 
future experiments searching for $\mu \to e$ conversion in nuclei, 
rare meson decays, and other processes.
\end{abstract}


\section{Introduction} \setcounter{equation}{0}

The masses of the six quarks and three charged leptons follow some
intriguing patterns.
At first sight, how heavy a fermion is depends crucially on which generation 
it belongs to.  
Fermions of the first generation are lighter by roughly 
two orders of magnitude than the corresponding fermions from the second generation,
which in turn are two orders
of magnitude lighter than the corresponding fermions from the third generation 
\cite{Yao:2006px}.

This pattern has motivated the study of models where the
couplings of the fermions to the electroweak symmetry breaking sector 
are linear in the standard model fermion fields, so that the dominant contributions to the fermion 
mass matrices have rank one ({\it i.e.}, only the third generation fermions 
have large masses). The masses for the second generation are then
induced at 1 loop \cite{Babu:1989fg} 
while first generation masses are further suppressed (attempts at deriving the 
electron mass from a loop involving the muon have a long history \cite{Georgi:1972hy}).

In an interesting scheme of this type \cite{Balakrishna:1988ks}, 
a pair of vectorlike fermions mix with the standard model quarks such that 
only the top and bottom quarks have tree-level masses. A charge 
$-1/3$ scalar then couples the third generation quarks to the other quarks, 
resulting in rank-two mass matrices at 1 loop and rank-three mass 
matrices at 2 loops. A similar scheme could be responsible 
for the lepton masses \cite{Balakrishna:1988xg,Balakrishna:1988bn}, although the current constraints 
on neutrino masses push the mass of the new particles introduced in \cite{Balakrishna:1988xg}
above $10^{14}$ GeV, while the constraints on lepton flavor violating processes 
in the model of \cite{Balakrishna:1988bn} render the muon and electron masses too 
small. 

However, the fermion masses follow more complicated patterns, as displayed in Figure 1.
Within the third generation, the $b$ quark and the $\tau$ lepton are almost two 
orders of magnitude lighter than the top quark. The charm quark, which belongs 
to the second generation, is only a few times lighter than the $b$ and $\tau$.
The other second generation fermions, namely the strange quark and the muon,
are lighter by an additional order of magnitude. 

\begin{figure}[t]\center
\psfrag{dd}[B]{}
\psfrag{uu}[B]{}
\psfrag{ll}[B]{}
\psfrag{GeV}[t]{\hspace{2cm} \parbox[t]{4.3cm}{\vspace*{-0.8cm}  Mass (GeV) } }
\psfrag{1}[t]{\hspace{2.5cm} \parbox[t]{3cm}{\vspace*{-0.3cm} \it $1$ } }
\psfrag{10}[t]{\hspace{2.3cm} \parbox[t]{3cm}{\vspace*{-0.3cm} \it $10$ } }
\psfrag{100}[t]{\hspace{2.2cm} \parbox[t]{3cm}{\vspace*{-0.3cm} \it $100$ } }
\psfrag{101}[t]{\hspace{2.2cm} \parbox[t]{3cm}{\vspace*{-0.3cm} \it $10^{-1}$ } }
\psfrag{102}[t]{\hspace{2.2cm} \parbox[t]{3cm}{\vspace*{-0.3cm} \it $10^{-2}$ } }
\psfrag{103}[t]{\hspace{2.2cm} \parbox[t]{3cm}{\vspace*{-0.3cm} \it $10^{-3}$ } }
\hspace*{0.9em}
\psfig{file=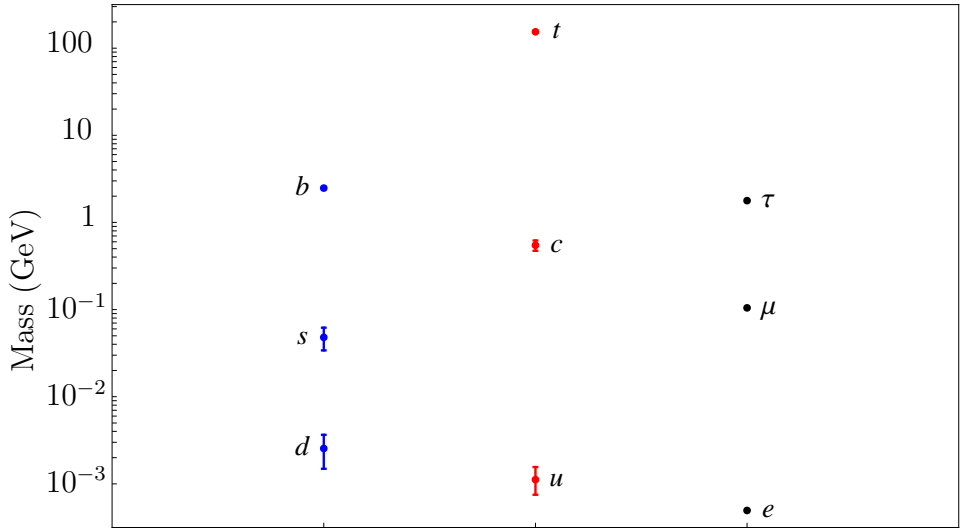,width=12.4cm,angle=0}
\vspace*{-5mm}
\caption{Quark and lepton masses at the 1 TeV scale, from Ref.~\cite{Xing:2007fb}.}
\label{fig:spectrum}
\end{figure}

Here we propose a mechanism for generating quark and lepton masses
based on the assumption that only the top quark mass arises at tree level.
In order to generate all the fermion masses, some new fields must couple the top 
quark to other standard model fermions.
One might expect that such a mechanism would require a large number of 
fields, and furthermore that all the masses would arise at one loop.
Remarkably, both these expectations turn out to be wrong due to the interplay of 
rank-one contributions to the mass matrices. 

Concretely, we first introduce one scalar field that couples the top quark 
to the leptons (Section 2). This leads to masses for the $\tau$, $\mu$ and the electron
at 1, 3 and 5 loops respectively. At the same time the charm and up quarks 
get masses at 2 and 4 loops respectively.
In Section 3 we demonstrate that the top quark may be the only standard model fermion 
that acquires mass at tree level even when there is no new quantum number that 
differentiates it from the other quarks\footnote{Alternatively,
the top quark may be the only fermion with a tree-level mass because of some 
symmetry acting on the standard model fermions. A related model, where 
an $S_3$ symmetry allows tree-level masses only for the 
top and bottom quarks, is given in Ref.~\cite{Babu:1990fr}.}.

In order to generate the remaining quark masses we introduce (Section 4) 
some additional fields that couple to the down-type quarks,
resulting in $b$, $s$ and $d$ masses at 1, 3 and 4 loops respectively.  
It turns out that these fields also contribute to the charm and up quark masses, 
and more importantly generate an electron mass at 4 loops.
It is remarkable that this realistic pattern of loop-induced masses arises without need for 
any flavor symmetry to differentiate the three generations\footnote{Other models of fermion
mass generation without flavor symmetries can be found, for example, 
in Refs.~\cite{Barr:1981wv, Balakrishna:1988ks}. }.
Furthermore, we show that the ensuing CKM matrix has elements consistent with experiment.  

Various phenomenological constraints, discussed in Section 5, 
require the masses of some of the new
particles to be substantially heavier than the electroweak scale. 
We envision that the gauge hierarchy problem  
is solved by supersymmetry or Higgs compositeness at the TeV scale. Although 
we do not explicitly embed our mechanism for fermion mass generation in a
more complete theory of that type, we do not expect that such an embedding would
encounter major hurdles. Note in particular that composite Higgs models based
on top condensation \cite{Miransky:1989ds, Bardeen:1989ds, Dobrescu:1997nm} lead 
automatically to a large top mass, providing the appropriate input for the mechanism 
presented here.  

The possibility that the mass of the top quark may be responsible for
all other fermion masses has been previously considered \cite{He:1989er,Babu:1990vx,Rattazzi:1990wu}.
Various obstacles \cite{Rattazzi:1990wu}, however, have prevented theories of this type from being 
realistic. In the model of Ref.~\cite{He:1989er} a weak-triplet VEV is essential for generating 
the lepton and first generation quark masses, such that the current constraints 
lead to an additional suppression of several orders of magnitude for all these
masses. In  the model of Ref.~\cite{Babu:1990vx}, if the mechanism is correctly continued all the way 
to the first generation, then the down quark turns out to be lighter than the up quark.

Our conclusions are collected in Section 6. In the Appendix we compute the 2-loop 
diagrams responsible for the charm mass.


\section{Loop-induced masses for charged leptons and up-type quarks}
\setcounter{equation}{0}\label{sec:lepton}

We assume that the electroweak symmetry is spontaneously broken by the vacuum expectation value 
of a Higgs doublet $H$, and that the only nonzero Yukawa coupling of $H$ to the standard model 
fermions is 
\be
- y_t \, \overline{u}_{R}^3  Q_L^3 \, H + {\rm H.c.} 
\ee
Here $Q_L^i$ is the quark doublet of the $i$th generation, $u_R^j$ is the up-type quark
singlet of the $j$th generation, and $y_t$ is a dimensionless parameter.
The above Yukawa coupling breaks explicitly the $[U(3)]^3$ global symmetry of the quark kinetic terms
down to a $U(1)_t \times U(2)_Q \times U(2)_u \times U(3)_d$ chiral symmetry, corresponding to unitary transformations acting on $Q_L^3$, $Q_L^{1,2}$, $u_R^{1,2}$ 
and the down-type quark singlets $d_R^j$, respectively. 
The top quark mass is generated at tree level ($m_t = y_t v_\inh >0$, where $v_\inh \approx 174$ GeV), 
while the other quarks and leptons remain massless so far.

Let us introduce a complex scalar field, $r$, which transforms under  
$SU(3)_c\times SU(2)_W\times U(1)_Y$ as (3, 2, +7/6). The normalization of hypercharge used here is
$Y = Q-T^3$, where $Q$ is the electric charge and $T^3$ is the diagonal $SU(2)_W$ generator. 
The $r$ component of $T^3=-1/2$ ($T^3=+1/2$) has electric charge +2/3 (+5/3).
The most general renormalizable interactions of  $r$ with standard model fermions
are given by
\be
\lambda_{ij} \, r\, \overline{u}_R^{\, i} L_L^j 
- \lambda^\prime_{ij} \, r \, \overline{Q}_L^{\, i} e_R^j 
+ {\rm H.c.} ~,
\label{r-tilde}
\ee
where $i,j=1,2,3$ label the generations, $L_L^j$ are the lepton doublets,
and $e_R^j$ are the $SU(2)_W$-singlet electrically-charged leptons.
The $\lambda_{ij}$ and $\lambda^\prime_{ij}$ coefficients are dimensionless complex parameters.

The interactions (\ref{r-tilde}) break explicitly the quark chiral symmetry down to
$U(1)_u \times U(3)_d$, and the lepton chiral symmetry $U(3)_L\times U(3)_e$ down 
to $U(1)_L$. Here the $U(1)_u$ charge is an overall phase of the $Q_L^i$ and $u_R^i$ fields,
while $U(1)_L$ charge is the lepton number. The conservation of these global charges implies that
$r$ carries baryon number +1/3 (same as $Q_L^j$) and lepton number +1 (same as $L_L^j$), so that 
it is a leptoquark.

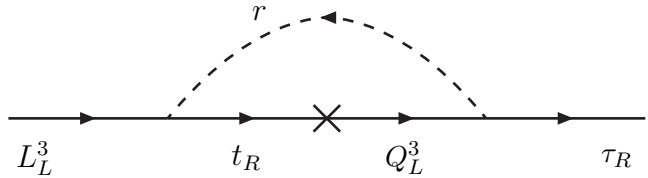
\begin{figure}[b!]
\vspace*{-0.5cm}
\begin{center} 
{
\unitlength=1 pt
\SetScale{1}\SetWidth{1}      
\normalsize    
{} \allowbreak
\begin{picture}(240,73)(0,-12)
\ArrowLine(0,0)(60,0)\ArrowLine(60,0)(120,0)\ArrowLine(120,0)(180,0)\ArrowLine(180,0)(240,0)
\DashCurve{(60,0)(105,35)(120,38)(135,35)(180,0)}{4}
\LongArrow(121,38)(119,38)
\Line(115,-5)(125,5)\Line(125,-5)(115,5)
\Text(10,-15)[c]{$L_L^3$}\Text(230,-15)[c]{\small $\tau_R$}
\Text(90,-15)[c]{$t_R$}\Text(150,-15)[c]{\small $Q_L^3$}
\Text(95,40)[c]{$r$}
\end{picture}
}
\end{center}
\vspace*{-0.4cm}
\caption{The 1-loop diagram responsible for the tau mass.
The $\times$ represents a top quark mass insertion.
}
\label{fig:tau}
\end{figure}

The breaking  of the chiral symmetries for the $Q_L$, $u_R$, $L_L$ and $e_R$ fields signals 
that all up-type quarks and electrically-charged leptons get masses at some loop level. 
Before computing the radiatively-induced masses, it is convenient to write 
the couplings (\ref{r-tilde}) in a basis where there are as many zeroes as possible. 
The most general form of $\lambda$ up to a $U(2)_u \times U(3)_L$ transformation is
\be
\lambda = \begin{pmatrix}
\lambda_{11} & \lambda_{12} & 0 \\
0 & \lambda_{22} & \lambda_{23} \\
0 & 0  & \lambda_{33} 
\end{pmatrix} ~,
\label{lambda}
\ee
where all $\lambda_{ij}$ are real and positive. 
Similarly, using the $U(2)_Q \times U(3)_e$ transformations, we can write
\be
\lambda^\prime = \begin{pmatrix}
\lambda^\prime_{11} & \lambda^\prime_{12} & 0 \\
0 & \lambda^\prime_{22} & \lambda^\prime_{23} \\
0 & 0  & \lambda^\prime_{33} 
\end{pmatrix} ~,
\ee
with  $\lambda_{ij}^\prime>0$.

Let us now identify the leading loop diagrams that communicate 
electroweak symmetry breaking from the top quark to the leptons and the charm quark. 
The $\tau$ mass is induced at 1 loop, as shown in Figure~\ref{fig:tau}, and is given by
\be
m_\tau \simeq \lambda_{33} \lambda^\prime_{33} \, m_t \, \epsilon^{(1)}_{r} ~,
\label{tau-mass}
\ee
where $\epsilon^{(1)}_{r}$ is the loop factor, which is logarithmically
divergent:
\be
\epsilon^{(1)}_{r} \simeq
\frac{N_c}{16\pi^2} \ln\left(\frac{\Lambda^2}{M_{r}^2}\right) ~.
\label{oneloop}
\ee
Here $N_c=3$ is the number of colors, $M_{r}$ is the mass of $r$, and $\Lambda$ is the cutoff 
scale where the quark (other than top) and lepton masses vanish.
For a cutoff $\Lambda \approx 10 M_{r}$ the loop factor is $\epsilon^{(1)}_{r} \approx 0.087$, and using 
the $m_\tau/m_t$ ratio at 1 TeV (see Figure 1) we find  
$\lambda_{33} \lambda^\prime_{33} \approx (0.36)^2$. 
In Section 3 we will present a simple renormalizable model where the cutoff 
$\Lambda$ is replaced by the mass of a new particle.

\begin{figure}[t!]
\vspace*{-0.5cm}
\begin{center} 
{
\unitlength=1 pt
\SetScale{1}\SetWidth{1}      
\normalsize    
{} \allowbreak
\begin{picture}(300,90)(0,-10)
\ArrowLine(0,0)(50,0)\ArrowLine(50,0)(100,0)\ArrowLine(100,0)(150,0)\ArrowLine(150,0)(200,0)
\ArrowLine(200,0)(250,0)\ArrowLine(250,0)(300,0)
\DashCurve{(100,0)(135,28)(150,31)(165,28)(200,0)}{4}
\LongArrow(151,31)(149,31)
\DashCurve{(50,0)(120,51)(150,55)(180,51)(250,0)}{4}
\LongArrow(149,55)(151,55)
\Line(145,-5)(155,5)\Line(155,-5)(145,5)
\Text(10,-15)[c]{$Q_L^2$}\Text(290,-15)[c]{\small $c_R$}
\Text(75,-15)[c]{$\tau_R$}\Text(225,-15)[c]{\small $L_L^3$}
\Text(125,-15)[c]{$Q_L^3$}\Text(175,-15)[c]{\small $t_R$}
\Text(134,32)[c]{$r$}\Text(126,60)[c]{$r$}
\end{picture}
}
\end{center}
\vspace*{-0.2cm}
\caption{Charm mass induced by the 2-loop ``rainbow'' diagram involving the $r$ scalar. 
}
\vspace*{0.6cm}
\label{fig:charm}
\end{figure}
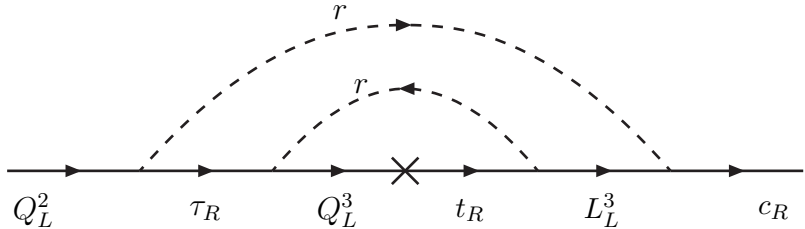

The charm quark mass is induced at two loops, through the ``rainbow'' diagram shown in 
Figure~\ref{fig:charm}. 
The entries in the up-type quark mass matrix from this type of diagrams are given by
\be
M_u[rr] =
\begin{pmatrix}
0 & 0 & 0 \\
0 & \lambda^\prime_{23} \lambda_{23} & \lambda^\prime_{33} \lambda_{23} \\
0 & \lambda^\prime_{23} \lambda_{33} & \lambda^\prime_{33} \lambda_{33} 
\end{pmatrix}
\lambda^\prime_{33} \lambda_{33} \, m_t \, \epsilon^{(2)}_{r}  ~,
\label{eq:charmone}
\ee
where $\epsilon^{(2)}_{r}$ is the 2-loop integral, and the corresponding term in the 
Lagrangian is $\overline{q}_R M_u q_L$.  
Approximating the inner loop in Figure~\ref{fig:charm} by Eq.~(\ref{oneloop}),
we find
\be
\epsilon^{(2)}_{r} \simeq \frac{1}{N_c} \left( \epsilon^{(1)}_{r} \right)^2 ~.
\label{epsilon-r2}
\ee
In Appendix A we show that this is a reasonable approximation.

In addition to  the ``rainbow'' diagram there are a few other 2-loop diagrams 
that connect a $Q_L^{2,3}$ external line to a $c_R$ or $t_R$ external line. These involve 
kinetic mixing between up-type quarks, and one can show that they do not
change the rank of the up-type mass matrix. As a result, they may be ignored 
in the computation of the charm mass.

Given that the tree-level top mass represents a large contribution to the 33 element 
of  the up-type quark mass matrix,
the charm mass is approximately given by the 22 element of  $M_u[rr]$:
\be
m_c \simeq \lambda^\prime_{23} \lambda_{23} \,  m_\tau \, \frac{\epsilon^{(1)}_{r}}{N_c} ~.
\ee
Assuming that there are no other contributions to the charm mass, the $m_c/m_\tau$ ratio at 1 TeV requires 
$\lambda_{23} \lambda^\prime_{23} \approx (3.3)^2$ for $\Lambda \approx 10 M_{r}$.  These Yukawa couplings are rather large, and one may worry that they do not remain perturbative up to the scale $\Lambda$.  However, in Section 4 it is shown that the sector responsible for the down-type quark masses actually leads to additional 2-loop contributions to $m_c$, so that the Yukawa couplings, $\myl_{23}$ and $\lp_{23}$, need not be that large.

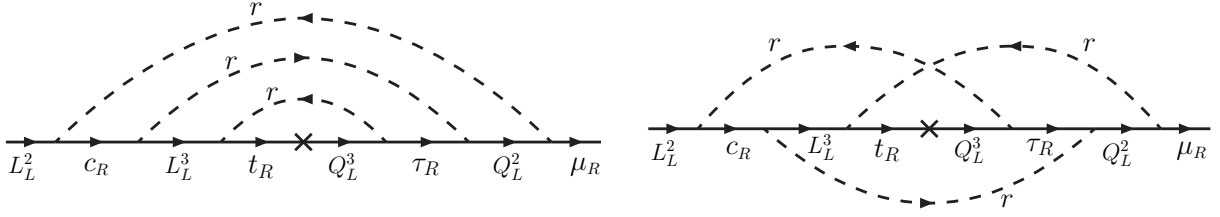
\begin{figure}[t]
\begin{center} 
\scalebox{.623}{
\unitlength=1 pt
\SetScale{1}\SetWidth{1.7}      
\normalsize\Large    
{} \allowbreak
\begin{picture}(500,113)(140,-20)
\ArrowLine(20,0)(50,0)\ArrowLine(50,0)(100,0)\ArrowLine(100,0)(150,0)\ArrowLine(150,0)(200,0)
\ArrowLine(200,0)(250,0)\ArrowLine(250,0)(300,0)\ArrowLine(300,0)(350,0)\ArrowLine(350,0)(380,0)
\DashCurve{(150,0)(185,24)(200,26)(215,24)(250,0)}{7}
\LongArrow(201,26)(199,26)
\DashCurve{(100,0)(170,46)(200,51)(230,46)(300,0)}{7}
\LongArrow(199,51)(201,51)
\DashCurve{(50,0)(160,69)(200,75)(240,69)(350,0)}{7}
\LongArrow(201,75)(199,75)
\Line(195,-5)(205,5)\Line(205,-5)(195,5)
\Text(30,-15)[c]{\large $L_L^2$}\Text(75,-15)[c]{$c_\inr$}
\Text(125,-15)[c]{\large $L_L^3$}\Text(175,-15)[c]{$t_R$}
\Text(225,-15)[c]{\large $Q^3_L$}\Text(275,-15)[c]{$\tau_R$}
\Text(325,-15)[c]{\large $Q^2_L$}\Text(372,-15)[c]{$\mu_\inr$}
\Text(182,29)[c]{$r$}\Text(157,48)[c]{$r$}\Text(172,81)[c]{$r$}
\put(380,8){
\ArrowLine(30,0)(60,0)\ArrowLine(60,0)(100,0)\ArrowLine(100,0)(150,0)\ArrowLine(150,0)(200,0)
\ArrowLine(200,0)(250,0)\ArrowLine(250,0)(300,0)\ArrowLine(300,0)(340,0)\ArrowLine(340,0)(370,0)
\DashCurve{(150,0)(220,46)(250,50)(280,46)(340,0)}{6.8}
\LongArrow(251,50)(249,50)
\DashCurve{(100,0)(170,-41)(200,-45)(230,-41)(300,0)}{7}
\LongArrow(199,-45)(201,-45)
\DashCurve{(60,0)(120,46)(150,50)(180,46)(250,0)}{6.8}
\LongArrow(151,50)(149,50)
\Line(195,-5)(205,5)\Line(205,-5)(195,5)
\Text(40,-15)[c]{\large $L_L^2$}\Text(85,-13)[c]{$c_\inr$}
\Text(135,-11)[c]{\large $L_L^3$}\Text(175,-13)[c]{$t_R$}
\Text(225,-13)[c]{\large $Q^3_L$}\Text(268,-11)[c]{$\tau_R$}
\Text(315,-15)[c]{\large $Q^2_L$}\Text(360,-13)[c]{$\mu_\inr$}
\Text(248,-44)[c]{$r$}\Text(107,51)[c]{$r$}\Text(298,51)[c]{$r$}
}
\end{picture}
}
\end{center}
\caption{Muon mass induced by the 3-loop rainbow and nonplanar diagrams involving the $r$ scalar.
In the nonplanar diagram there is no  intersection of the upper two $r$ lines.
}
\label{fig:mu}
\vspace*{0.6cm}
\end{figure}

Now that the charm quark has a mass, it will generate masses for the muon and up quark in the same way that 
the top mass lead to tau and charm masses. More precisely,
the leading contributions to the muon mass arise from 3-loops diagrams involving one top-mass insertion. 
There are only two nonzero diagrams: 
a rainbow and a nonplanar diagram, shown in Figure~\ref{fig:mu}. As in the case of the charm mass,
diagrams involving kinetic mixing on internal or external fermion lines may be ignored because
they do not change the rank of the matrix (a more transparent argument is given in Section 3).  
The charged-lepton mass matrix gets the following 
contributions from diagrams involving three $r$ lines:
\be
M_e[rrr] =
\begin{pmatrix}
\; 0 & 0 & 0 \\[1mm]
\; 0 & \lambda^\prime_{22}\lambda^\prime_{23}  \lambda_{23}\lambda_{22} 
& \lambda^\prime_{22}\lambda^\prime_{23} \left[(\lambda_{23})^2 + (\lambda_{33})^2\right] \\[3mm]
\; 0 & \;  \left[(\lambda^{\prime}_{23})^2\! +\! (\lambda^{\prime}_{33})^2\right]\lambda_{23}\lambda_{22}  
\; & \left[(\lambda^{\prime}_{23})^2\! +\! (\lambda^{\prime}_{33})^2\right]\left[(\lambda_{23})^2\! +\! (\lambda_{33})^2\right]
\end{pmatrix}
\lambda^\prime_{33} \lambda_{33} \, m_t \, \epsilon^{(3)}_{r}~. 
\ee
In the large $N_c$ limit the nonplanar diagram in Figure~\ref{fig:mu} is subleading to 
the rainbow diagram.  In addition, the nonplanar diagram involves fewer factors of 
$\ln (\Lambda^2/M_r^2)$ 
(the 2-loop computations given in the Appendix include an 
explicit example of how fewer logarithmic factors arise in a non-rainbow diagram).
Due to the combination of $N_c$ and logarithmic factor suppression, we expect that the rainbow diagram dominates.  So, the 3-loop factor, $\epsilon^{(3)}_{r}$, is given by
\be
\epsilon^{(3)}_{r} \simeq \frac{1}{N_c} \left( \epsilon^{(1)}_{r} \right)^3  ~.
\ee
The 33 element of  the charged-lepton mass matrix is dominated by the 
the 1-loop tau mass from Eq.~(\ref{tau-mass}), so that
the muon mass is approximately given by the 22 element of  $M_e[rrr]$:
\be
m_\mu \simeq  \lambda^\prime_{22} \lambda_{22} \, m_c \, \epsilon^{(1)}_{r} ~.
\ee
The $m_\mu/m_c$ ratio at 1 TeV requires $\lambda_{22} \lambda^\prime_{22} \approx (1.5)^2$. 

The up-quark mass is generated at 4 loops. There are five diagrams, each involving four $r$ 
lines. All these diagrams have 8 vertices, proportional to $\myl_{12}$, $\myl_{22}$,
 $\myl_{23}$, $\myl_{33}$, $\lp_{33}$,  $\lp_{32}$,  $\lp_{22}$, and  $\lp_{21}$, respectively.
The only difference between the diagrams comes from the way the four outgoing  $r$ lines
are contracted with the four incoming  $r$ lines. If we label the above vertices by
1, 2, ... , 8, the pairing of $r$ lines in the rainbow diagram is 18-27-36-45.
In the large $N_c$ limit, the rainbow diagram dominates, being of order $N_c^2$. 
However, there are three other diagrams (18-25-36-47, 16-27-38-45, 14-27-36-58) of order $N_c$ 
which cannot be neglected for $N_c=3$. It is likely though that some of these
diagrams have fewer factors of $\ln (\Lambda^2/M_r^2)$ than the rainbow diagram.
The remaining diagram (16-25-38-47) does not depend on $N_c$.
Thus, even though all these five diagrams have the same sign and add constructively, 
their sum may be reasonably well approximated by the rainbow diagram. 
The 4-loop contributions to the up-quark mass matrix take the form
\be
M_u[rrrr]_{ij}= \left( \sum_{a,b,c,d} \myl_{ia} \myl_{ba} \myl_{b3} \lp_{c3} \lp_{cd} \lp_{jd} \right)
\myl_{33} \lp_{33} \, m_t \, \epsilon_{r}^{(4)} ~,
\ee
where the 4-loop factor is expected to be of order
\be
\epsilon_{r}^{(4)} \sim  \frac{1}{N_c^2} \left( \epsilon_{r}^{(1)}\right)^4 ~.
\label{4-loop}
\ee
The up quark mass is given approximately by the 11 entry of $M_u[rrrr]$:
\be
m_u\approx \lp_{12} \myl_{12} \, m_\mu \frac{\epsilon_{r}^{(1)}}{N_c}  ~.
\ee
In the absence of contributions to the up mass from a different sector (see Section 4), 
the $m_u/m_\mu$ ratio at 1 TeV requires $\lambda_{12} \lambda^\prime_{12} \approx (0.6)^2$, where we 
ignored the order-one uncertainty introduced by Eq.~(\ref{4-loop}).

Finally, the electron mass arises at 5 loops. There are two diagrams at order
$N_c^3$, eight diagrams at order $N_c^2$ and eleven diagrams at order $N_c$, 
all involving five $r$ lines. Ignoring the uncertainty associated with the 
sum of these diagrams, we estimate 
\be
m_e \sim \lp_{11} \myl_{11} \, m_u O(\epsilon_{r}^{(1)}) ~.
\label{electron}
\ee
The $m_u/m_e$ ratio at 1 TeV requires $\lambda_{11} \lambda^\prime_{11} \approx (2.3)^2$.  
As in the case of the $u$ or $\mu$ mass, all the diagrams contributing to the 
electron mass have the same sign. In the hypothetical case where all the loop integrals 
for non-rainbow diagrams have the same size as the 
rainbow one, the estimate for 
$\myl_{11}\lp_{11}$ is smaller by a factor of $\sim (2.4)^2$.

The constraints on various processes induced by leptoquark exchange set limits
on $M_r$ far above 1 TeV (see section 5), so that the renormalization group evolution 
changes the quark and lepton masses at the scale $M_r$ compared to the values shown in Figure 1.
The most notable effect is that the quark masses decrease faster than the lepton masses 
when the scale where they are evaluated increases \cite{Xing:2007fb}. 
We have not taken this effect into account in this paper, because
the field content above the TeV scale is not uniquely determined.

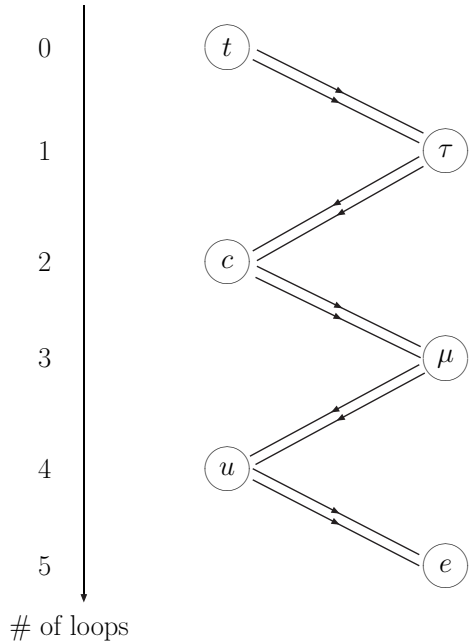
\begin{figure}[b!]
\begin{center} 
\scalebox{.523}{
\unitlength=1 pt
\SetScale{1}\SetWidth{1}      
\normalsize\Large    \LARGE
{} \allowbreak
\begin{picture}(400,460)(-30,-115)
\put(97,300){\circle{30}}\put(255,225){\circle{30}}
\put(97,145){\circle{30}}\put(255,75){\circle{30}}
\put(97,-5){\circle{30}}\put(255,-75){\circle{30}}
\ArrowLine(115,290)(235,230)\ArrowLine(118,297)(238,237)
\ArrowLine(235,220)(115,152)\ArrowLine(238,213)(118,145)
\ArrowLine(118,141)(236,83)\ArrowLine(117,133)(236,75)
\ArrowLine(235,70)(113,4)\ArrowLine(239,64)(117,-2)
\ArrowLine(115,-5)(235,-67)\ArrowLine(115,-13)(235,-75)
\Text(97,300)[c]{$t$}\Text(255,225)[c]{$\tau$}
\Text(97,145)[c]{$c$}\Text(255,75)[c]{$\mu$}
\Text(97,-5)[c]{$u$}\Text(255,-75)[c]{$e$}
\Text(39,-120)[c]{\parbox{7cm}{\# of loops}}
\Text(-35,-75)[c]{5}\Text(-35,-5)[c]{4}
\Text(-35,75)[c]{3}\Text(-35,145)[c]{2}
\Text(-35,225)[c]{1}\Text(-35,300)[c]{0}
\multiput(-7,330)(.85,0){2}{\vector(0,-1){430}}
\LongArrow(-6.7,-95)(-6.7,-100)
\end{picture}
}
\end{center}
\caption{Loop-level where masses for charged leptons and up-type quarks are generated. 
Each line connecting a pair of fermions
indicate Yukawa interactions with $r$. }
\label{fig:map}
\end{figure}

In summary, if the only source of mass for the up-type quarks and charged leptons is loops 
involving $r$, then the observed fermion masses, which span almost six orders of magnitude,
may be obtained with values for the Yukawa couplings
of $r$ ranging between 0.36 and 3.3. Furthermore, the observed mass ordering 
$m_t > m_\tau > m_c > m_\mu > m_u > m_e$ is correctly reproduced by the number of loops 
required for generating each of these masses.  Figure~\ref{fig:map} depicts schematically 
how the masses for these fermions are generated. It is interesting to compare this figure 
with the fermion mass spectrum shown in Figure~\ref{fig:spectrum}, keeping in mind that 
the mass decreases exponentially as the number of loops increases linearly. 

In generating lepton masses from the top quark mass, a field with the quantum numbers of a 
leptoquark is generically necessary.  
There is however an alternative to the leptoquark ($r$) included here: a scalar $\tilde{d}$ transforming as $(\overline{3},1,+1/3)$
under $SU(3)_c\times SU(2)_W\times U(1)_Y$. The $\tilde{d}$ leptoquark has the quantum numbers of a right-handed 
down-type squark (note that in supersymmetric models with R-parity violation the squarks 
may have leptoquark couplings).
The most general gauge invariant Yukawa couplings to the standard model fermions are 
\be
\kappa_d \, \tilde{d}\, \overline{Q}_L^{\, c}\, L_L + \, \kappa^\prime_d \, \tilde{d}\, \overline{u}_R^{\, c}\, e_R + {\rm H.c.} ~,
\ee
where the flavor structure of the $\kappa_d$ and $\kappa^\prime_d$ couplings is the same as in Eq.~(\ref{lambda}).
The analysis carried out for $r$ also applies to $\tilde{d}$: the above couplings break the chiral symmetries and  
lead to up-type quark and lepton masses, with the same loop counting. 
We will not  discuss further the $\tilde{d}$ leptoquark in this paper.

\bigskip

\section{Renormalizable UV completion}\setcounter{equation}{0}

In the previous section we have assumed that the Higgs doublet couples only to the 
top quark at tree level. In this section we are going to justify this assumption 
by introducing a new symmetry acting on the Higgs sector (but not on the standard fermions) in a renormalizable model.


We introduce a symmetry, $G_H$, under which the Higgs doublet is charged while all standard model fermions are singlets.  This forbids any dimension-4 couplings of the Higgs doublet to standard model fermions.  The new symmetry is broken by the VEV of a scalar field $\phi$ which is a singlet under $SU(3)_c\times SU(2)_W\times U(1)_Y$.
At this stage the chiral symmetries of the standard model are unbroken.  

Examples of this symmetry could be a gauge or global $U(1)_H$, or a discrete subgroup thereof.  
We will consider, for concreteness, a global $U(1)_H$.  As we will see, our minimal model 
would have to be extended if $G_H$ is gauged since its fermion content 
is anomalous.  For the case of a global $U(1)_H$ considered here this anomaly implies that 
the Goldstone boson has a small mass.  However, this mass is not sufficient for 
$\langle \phi \rangle \!\! \lae \! 10^7$ TeV
to avoid constraints from star cooling, and some additional small explicit 
breaking of $U(1)_H$ must be included to increase the mass of the would-be Goldstone boson.  
If $G_H$ were discrete, then these constraints would be avoided, but instead one 
would need to ensure that the associated domain walls are cosmologically allowed.  
The solution to these problems should not affect the predictions we make here.

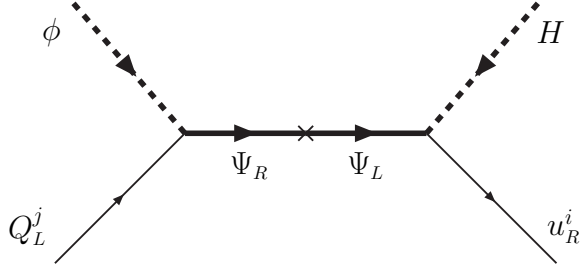
\begin{figure}[t]
\begin{center} 
\scalebox{.7}{
\unitlength=1 pt
\SetScale{1}\SetWidth{3}      
\normalsize\Large    
{} \allowbreak
\begin{picture}(300,160)(-10,10)
\SetWidth{1} 
\ArrowLine(0,0)(70,70)\ArrowLine(200,70)(270,0)
\Text(-15,20)[c]{$Q_\inl^j$}\Text(275,20)[c]{$u^i_\inr$}
\Line(139,66)(131,74)\Line(131,66)(139,74)
\SetWidth{3} 
\ArrowLine(70,70)(135,70)\ArrowLine(135,70)(200,70)
\DashArrowLine(10,140)(70,70){4}\DashArrowLine(260,140)(200,70){4}
\Text(-2,125)[c]{$\phi$}\Text(268,125)[c]{$H$}
\Text(105,53)[c]{$\Psi_\inr$}\Text(168,53)[c]{$\Psi_\inl$}
%
\end{picture}
}
\end{center}
\caption{Top mass from an extended Higgs sector.
}
\label{fig:psi}
\end{figure}

We introduce a vectorlike fermion, $\Psi$, transforming as $Q_L$ under 
$SU(3)_c\times SU(2)_W\times U(1)_Y$, which carries $U(1)_H$ charge $-1$.
Then the most general Yukawa couplings of $\phi$ and $H$ are given by
$H \overline{u}_R^i \Psi_L$ and $\phi \overline{\Psi}_R Q_L^j$.
Without loss of generality we can use the chiral transformations to rewrite these 
Yukawa couplings as
\be
-y_\inh H \, \overline{u}_R^3 \Psi_L - y_\phi \phi \, \overline{\Psi}_R Q_L^3 + {\rm H.c.}
\ee
where $y_H$ and $y_\phi$ are real positive parameters.

Integrating out the heavy $\Psi$ fermion (see Figure~\ref{fig:psi}) leads to 
a dimension-5 operator $\phi  H \, \overline{u}_R^3 \, Q_L^3$.
Replacing the $\phi$ scalar by its VEV then leads to an
effective top Yukawa coupling. The mass
scale that suppresses this operator is 
given by the $\Psi$ mass $M_\Psi$ if  $M_\Psi \gg y_\phi \langle \phi \rangle$. 
More generally, there is a $2\times 2$ mass matrix for $\Psi$ and top, whose lighest 
eigenvalue is the physical top mass. 
This is similar to the top-seesaw theory of Higgs compositeness
\cite{Dobrescu:1997nm}.
For $y_\inh  v_\inh \ll y_\phi \langle \phi \rangle$ and  $y_\inh v_\inh \ll M_\Psi$,
\be
m_t \approx y_\inh v_\inh \left[ 1 + \left(\frac{M_\Psi}{y_\phi \langle \phi \rangle}\right)^{\!2} \,
\right]^{-1/2} ~.
\label{top-mass}
\ee
A remarkable thing has happened: 
only the top quark 
acquires mass at tree level even when no symmetry 
differentiates it from other standard model fermions!

\begin{figure}[t]
\begin{center} 
{
\unitlength=1 pt
\SetScale{1}\SetWidth{1}      
\normalsize    
{} \allowbreak
\begin{picture}(340,100)(-10,-45)
\scalebox{0.95}{
\ArrowLine(20,0)(60,0)\ArrowLine(60,0)(120,-20)
\ArrowLine(120,-20)(180,-20)\ArrowLine(180,-20)(240,-20)
\ArrowLine(240,-20)(300,0)\ArrowLine(300,0)(340,0)
\DashArrowLine(120,-20)(110,-62){4}\DashArrowLine(240,-20)(250,-62){4}
\DashCurve{(60,0)(150,35)(180,38)(210,35)(300,0)}{4}
\LongArrow(181,38)(179,38)
\Line(185,-15)(175,-25)\Line(175,-15)(185,-25)
\Text(30,-15)[c]{\small $L_L^3$}\Text(330,-15)[c]{$\tau_\inr$}
\Text(90,-25)[c]{$t_\inr$}\Text(273,-25)[c]{\small $Q_L^3$}
\Text(150,-35)[c]{\small $\Psi_L$}\Text(210,-35)[c]{\small $\Psi_R$}
\Text(165,45)[c]{$r$}
\Text(98,-60)[c]{$H$}\Text(261,-60)[c]{\small $\phi$}
}
\end{picture}
}
\end{center}
\caption{Effective operator responsible for the tau mass.
}
\label{fig:finite}
\end{figure}
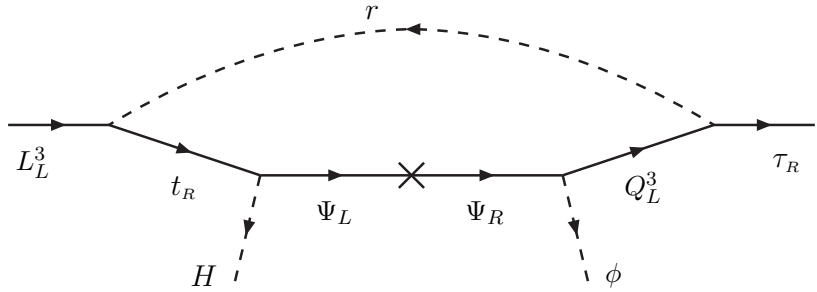

We can now repeat the analysis of Section 2 except instead of giving
logarithmically divergent contributions to the fermion masses, the loops
will generate {\it finite} coefficients to dimension-5 operators involving
$\phi$, $H$ and a standard model fermion pair. For example, the 
coefficient of the $\phi H \, \overline{\tau}_R L_L^3$ operator 
arises from the 1-loop diagram shown in Figure~\ref{fig:finite},
and is given by
$y_\inh y_\phi \lambda_{33} \lambda^\prime_{33} \, N_c /M_\Psi$ 
times a finite integral:
\bear
 I_1(M_\Psi,M_r) & = & M_\Psi^2 \int\frac{d^4 k}{(2\pi)^4}\; \frac{i}{k^2\left(k^2-M_\Psi^2\right)
\left(k^2-M_r^2\right)} \nonumber \\ [3mm] 
& = & \frac{1}{16\pi^2} \, \frac{M_\Psi^2}{M_\Psi^2-M_r^2}
\ln\left(\frac{M_\Psi^2}{M_r^2}\right) ~.
\eear
Replacing $H$ and $\phi$ by their VEVs, and using Eq.~(\ref{top-mass}), we find the tau mass:
\be
m_\tau \simeq \lambda_{33} \lambda^\prime_{33} \, 
N_c m_t  \left[ 1 + \left(\frac{y_\phi \langle \phi \rangle}{M_\Psi}\right)^{\!2} \, \right]^{1/2}
 I_1(M_\Psi,M_r) ~.
\ee
Comparing this result with Eq.~(\ref{tau-mass})
for $M_\Psi^2 \gg (y_\phi \langle \phi \rangle )^2$ and $M_\Psi \gg M_r$ shows that the cutoff 
scale used in Section 2 may be identified with the $\Psi$ mass:
$\Lambda \simeq M_\Psi $.  For convenience we take this limit in what follows.

The UV completion discussed here results in only the top quark acquiring a tree-level coupling to the Higgs doublets,
and implies that all the Yukawa couplings come from dimension-5 operators, of the form $\phi\, H\, \overline{\psi}_R\, \psi_L$.  The lack of a renormalizable counterterm means that all fermion mass terms are finite.  This justifies ignoring diagrams which involve kinetic mixing on internal or external lines, as we did earlier, because they contain fewer propagators in the loops and would lead to a mass with logarithmic dependence on the cutoff scale, which is forbidden by $U(1)_H$.  Since kinetic mixing does not change the rank of the mass matrices, its presence in diagrams cannot lead to loop generated masses and it can be ignored.

\section{Loop-induced down-type quark masses}\setcounter{equation}{0}
\label{sec:downtype}

With the fields and interactions introduced in the previous sections 
the chiral symmetry of the Lagrangian is $U(3)_d \times U(1)_u \times U(1)_L$.  
In order to generate masses for the $b$, $s$ and $d$ quarks, some 
new interactions must break the $U(3)_d$ symmetry by coupling the 
right-handed down-type quarks to fields involved in electroweak 
symmetry breaking. There are several possible interactions  
of this type. In this Section we focus for definiteness on a particular 
set of interactions.

\subsection{$b$-quark mass}

Let us introduce a pair of scalar fields,
$\he$ and $\hep$, which transform under  
$SU(3)_c\times SU(2)_W\times U(1)_Y$ as $(8,2,\pm 1/2)$,
and carry global $U(1)_H$ charge $+1$.
At the renormalizable level, the most general couplings of $\he$ and $\hep$ to
fermions are
\be
\kappa_{i} \, \Phi_8 \; \overline{u}^{i}_R \Psi_L 
+ \kappa^\prime \, \Phi_8^\prime \; \overline{d}^{3}_R \Psi_L + \mathrm{H.c.} ~,
\label{eq:h8}
\ee
where $\kappa_3$ and $\kappa^\prime$ are positive parameters (up to a phase redefinition
of $\he$ and $\hep$),
while $\kappa_1$ and $\kappa_2$ are complex dimensionless parameters.
Just as the Higgs doublet couples to only one linear combination of up-type 
quarks, $\hep$ couples to only one linear combination of down-type quarks, 
which defines the bottom quark. The above interactions break 
explicitly the $U(3)_d \times U(1)_u$ chiral symmetry down to $U(2)_d$, so that
they induce a mass for the $b$ quark and not for the $s$ and $d$ quarks.

Besides the usual quartic couplings for the scalars, the
\be
c \, \Phi_8 \Phi_8^\prime \phi \phi 
+ c^{\,\prime} \, H \Phi_8^\dagger r^\dagger r 
+ c^{\,\prime\prime} \left(\Phi_8\,H^\dagger \right)^2 + \mathrm{H.c.}
\label{quartic}
\ee
quartic couplings are allowed by all symmetries (the last coupling will not be 
used in the generation of fermion masses). The coefficients $c^\prime$ and 
$c^{\prime\prime}$ are complex numbers, while $c$ is real and positive (its phase is absorbed 
by a redefinition of the $\Phi_8^\prime$ field, which in turn requires the same phase
to be absorbed into  $d_R^3$ in order to keep $\kappa^\prime$ real).

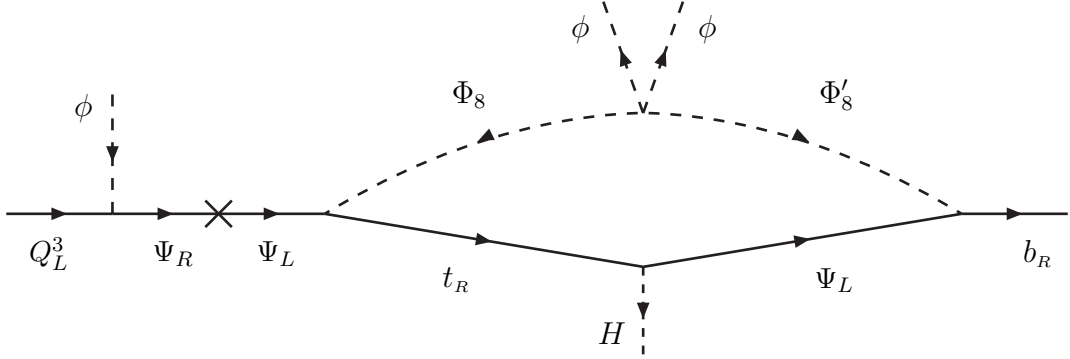
\begin{figure}[t!]
\vspace*{1.2cm}
\begin{center} 
{
\unitlength=1 pt
\SetScale{1}\SetWidth{1}      
\normalsize    
{} \allowbreak
\begin{picture}(350,120)(-30,-80)
\ArrowLine(-60,0)(-20,0)\ArrowLine(-20,0)(20,0)\ArrowLine(20,0)(60,0)
\ArrowLine(60,0)(180,-20)\ArrowLine(180,-20)(300,0)\ArrowLine(300,0)(340,0)
\DashArrowLine(180,-20)(180,-53){3.5}
\DashArrowLine(180,38)(195,80){4}\DashArrowLine(180,38)(165,80){4}
\DashArrowLine(-20,45)(-20,0){4}
\DashCurve{(60,0)(150,35)(180,38)(210,35)(300,0)}{4}
\LongArrow(121,29)(119,28)\LongArrow(239,29)(241,28)
\Line(25,-5)(15,5)\Line(15,-5)(25,5)
\Text(-44,-15)[c]{\small $Q_L^3$}\Text(3,-15)[c]{\small $\Psi_R$}
\Text(42,-15)[c]{\small $\Psi_L$}\Text(330,-15)[c]{$b_\inr$}
\Text(110,-25)[c]{$t_\inr$}\Text(253,-25)[c]{\small $\Psi_L$}
\Text(169,-45)[c]{$H$} \Text(-31,40)[c]{$\phi$}
\Text(157,70)[c]{$\phi$}\Text(205,70)[c]{$\phi$}
\Text(115,45)[c]{$\Phi_8$}\Text(254,45)[c]{$\Phi_8^\prime$}
\end{picture}
}
\end{center}
\vspace*{-1.1cm}
\caption{$b$-quark mass generated at one loop. The gauge-singlet scalar has a VEV   
$\langle\phi\rangle \sim M_\Psi$. }
\label{fig:bmass}
\end{figure}

The $b$ quark acquires a positive mass at one loop from the diagram shown in 
Figure~\ref{fig:bmass}: 
\be
m_b=  \kappa_3 \kappa^\prime c \,
m_t\, \langle\phi\rangle^2 
\, N_c \, \widetilde{I}_1(M_\Psi,M_8,M_{8'}) ~,
\ee
where $M_8$ and $M_{8'}$ are the masses of the $\he$ and $\hep$ scalars, and
\bear
\label{eq:bloop}
\widetilde{I}_1(M_\Psi,M_{8},M_{8'}) & \equiv & \int\frac{d^4 k}{(2\pi)^4}\; \frac{i}{(k^2-M_\Psi^2)\left(k^2-M_8^2\right)
\left(k^2-M_{8'}^2\right)}  \\ [3mm] 
& = & 
\frac{M_{8'}^2\, M_8^2 \ln \left(  M_{8'}/M_8\right)
+ M_{\Psi}^2\, M_{8'}^2 \ln \left( M_{\Psi}/M_{8'} \right)
+ M_{8}^2\, M_{\Psi}^2 \ln \left(M_{8}/M_{\Psi}\right)}
{8\pi^2\,(M_{8'}^2-M_{8}^2)\,(M_{\Psi}^2-M_{8}^2)\,(M_{\Psi}^2-M_{8'}^2)} ~. \nonumber
\eear
Taking the limit $M_8 \ll M_\Psi,\, M_{8'}$, and to leading order in $(M_{8'}/M_{\Psi})^2$,
the integral of (\ref{eq:bloop}) becomes
\be
\widetilde{I}_1(M_\Psi,M_{8},M_{8'}) \approx \frac{1}{16\pi^2 \, M_\Psi^2} \ln \left( \frac{M_{\Psi}^2}{M_8^{\prime 2}}\right) ~.
\ee
As before, this mass is a finite effect with the logarithm being cutoff by the mass of 
the massive fermion.  Working in this regime and assuming that 
$\langle\phi\rangle \approx M_\Psi$ and $y_\phi^2\ll 1$ 
we find that the correct $b$-quark mass requires $\kappa_3\, \kappa' c \approx (0.6)^3$, for 
$M_{\Psi}/M_{8'} \sim O(10)$.

In addition to the masses generated in section~\ref{sec:lepton} when $r$ was 
integrated out, 
there are potentially important contributions to the charm, up and electron 
masses from the fields introduced in this section.

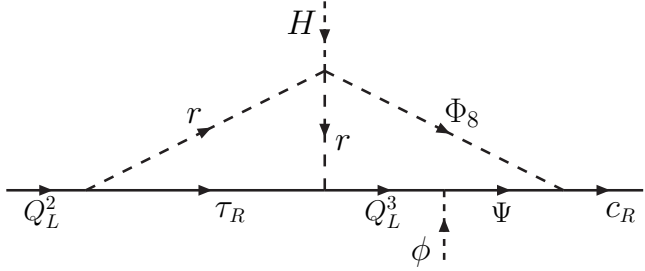
\begin{figure}[t]
\begin{center} 
\scalebox{0.75}{
\unitlength=1 pt
\SetScale{1}\SetWidth{1.3}      
\normalsize   \Large 
{} \allowbreak
\begin{picture}(340,100)(10,-13)
\ArrowLine(20,0)(60,0)\ArrowLine(60,0)(180,0)
\ArrowLine(180,0)(240,0)\ArrowLine(240,0)(300,0)\ArrowLine(300,0)(340,0)
\DashArrowLine(180,60)(180,0){5.5}\DashArrowLine(180,95)(180,60){4}
\DashArrowLine(60,0)(180,60){5.5}
\DashArrowLine(180,60)(300,0){5.5}
\DashArrowLine(240,-35)(240,0){4}
\Text(38,-11)[c]{\large $Q_L^2$}
\Text(133,-11)[c]{$\tau_\inr$}
\Text(210,-11)[c]{\large $Q_L^3$}\Text(270,-11)[c]{\large $\Psi$}
\Text(330,-11)[c]{$c_\inr$}
\Text(190,24)[c]{$r$}\Text(169,85)[c]{$H$}  
\Text(115,38)[c]{$r$}\Text(250,39)[c]{$\Phi_8$}
\Text(229,-31)[c]{$\phi$}  
\end{picture}
}
\end{center}
\caption{Charm-quark mass induced at 2 loops by the $\Phi_8$ interactions. 
}
\label{fig:charm-mass}
\end{figure}

The charm-quark mass receives an additional two-loop contribution 
from the diagram shown in Figure~\ref{fig:charm-mass}.
This is a direct contribution from the Higgs VEV, unlike the 
two-loop contribution to the charm mass discussed in Section 2, which first required the tau 
to get a mass. However, as before, this new contribution does involve a top-quark internal line. 
The new two-loop contributions to the mass matrix of the up-type quarks are given by
\be
\label{eq:charmtwo}
M_u[\Phi_8 r]=\begin{pmatrix}
0 & \ \kappa_1 \lp_{23} \ & \kappa_1 \lp_{33}\\
0 & \kappa_2 \lp_{23} & \kappa_2 \lp_{33} \\
0 & \kappa_3 \lp_{23} & \kappa_3 \lp_{33}
\end{pmatrix} 
\lp_{33} c' \, \frac{y_\phi\, \langle \phi\rangle\, v_\inh}{M_\Psi} \, \epsilon_\Phi^{(2)} ~.
\ee
Here $\epsilon_\Phi^{(2)}$ is a 2-loop integral computed in the Appendix [see Eq.~(\ref{eq:tent})],
which is parametrically smaller than the $\epsilon_r^{(2)}$ integral of Eq.(\ref{epsilon-r2}) by a 
logarithmic factor. Nevertheless, the above contribution to the charm mass is not 
suppressed by the small product of couplings $\lambda_{33}\lambda_{33}^\prime$, 
and therefore it may be comparable to or even larger than the rainbow diagram of 
Figure~\ref{fig:charm}.
Consequently, $\lambda_{23}$ and $\lambda'_{23}$ may be substantially smaller
than the values determined in Section~\ref{sec:lepton}.


\begin{figure}[t]
\begin{center} 
\scalebox{0.96}{
\unitlength=1 pt
\SetScale{1}\SetWidth{1.2}      
\normalsize   
{} \allowbreak
\begin{picture}(480,110)(10,-23)
\ArrowLine(0,0)(20,0)\ArrowLine(20,0)(50,0)\ArrowLine(50,0)(80,0)\ArrowLine(80,0)(110,0)
\ArrowLine(110,0)(140,0)\ArrowLine(140,0)(170,0)\ArrowLine(170,0)(200,0)\ArrowLine(200,0)(220,0)
\DashArrowLine(50,0)(80,50){5.5}\DashArrowLine(80,50)(107,0){4}\DashArrowLine(200,0)(80,50){5.5}
\DashCurve{(20,0)(60,-31)(80,-34)(100,-31)(140,0)}{5.5}
\LongArrow(81,-34)(79,-34)
\DashCurve{(80,0)(115,-30)(125,-32)(135,-30)(170,0)}{5.5}
\LongArrow(124,-32)(126,-32)
\DashArrowLine(80,50)(80,74){3}
\Text(8,-11)[c]{\small $L_L^1$}\Text(41,9)[c]{$u_\inr$}\Text(63,-11)[c]{\small $Q_L^3$}
\Text(99,-11)[c]{$\tau_\inr$}\Text(116,11)[c]{\small $Q_L^2$}\Text(148,-9)[c]{$\mu_\inr$}
\Text(183,-11)[c]{\small $Q_L^1$}\Text(213,-11)[c]{$e_\inr$}
\Text(105,24)[c]{$r$}\Text(156,29)[c]{$r$}
\Text(50,-34)[c]{$r$}\Text(146,-34)[c]{$r$}
\Text(93,68)[c]{$H$}\Text(55,32)[c]{$\Phi_8$}\Text(51,-2)[c]{\Large $\bullet$}
\put(250,0){
\ArrowLine(0,0)(20,0)\ArrowLine(20,0)(50,0)\ArrowLine(50,0)(80,0)\ArrowLine(80,0)(110,0)
\ArrowLine(110,0)(140,0)\ArrowLine(140,0)(170,0)\ArrowLine(170,0)(200,0)\ArrowLine(200,0)(220,0)
\DashArrowLine(80,50)(20,0){5.5}\DashArrowLine(50,0)(80,50){4}\DashArrowLine(140,0)(80,50){5.5}
\DashCurve{(78,0)(115,-27)(125,-29)(135,-27)(171,0)}{5.5}
\LongArrow(124,-29)(126,-29)
\DashCurve{(108,0)(145,-27)(155,-29)(165,-27)(199,0)}{5.5}
\LongArrow(156,-29)(154,-29)
\DashArrowLine(80,50)(80,74){3}
\Text(8,-11)[c]{\small $L_L^1$}\Text(41,-11)[c]{$u_\inr$}\Text(66,-11)[c]{\small $Q_L^3$}
\Text(101,-10)[c]{$\tau_\inr$}\Text(131,-11)[c]{\small $Q_L^2$}\Text(154,8)[c]{$\mu_\inr$}
\Text(183,11)[c]{\small $Q_L^1$}\Text(213,-11)[c]{$e_\inr$}
\Text(42,28)[c]{$r$}\Text(123,27)[c]{$r$}
\Text(110,-33)[c]{$r$}\Text(171,-33)[c]{$r$}
\Text(91,68)[c]{$H$}\Text(78,21)[c]{$\Phi_8$}\Text(50,-1)[c]{\Large $\bullet$}
}
\end{picture}
}
\end{center}
\caption{Electron mass induced at 4 loops. The diagrams are nonplanar (the lower two $r$ 
lines do not intersect).
The $\bullet$ indicates a vertex obtained by 
integrating out the $\Psi$ fermion.}
\label{fig:electron-mass}
\end{figure}
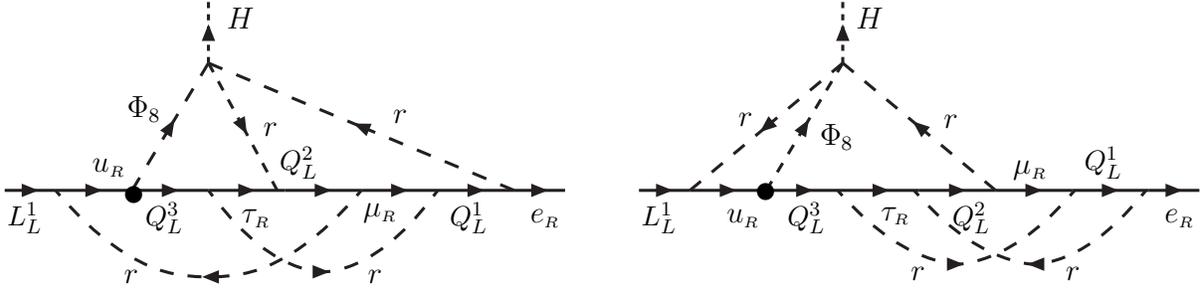

The up-quark mass also gets additional contributions, at 4 loops, from diagrams that 
involve one $H\he^\dagger r^\dagger r$ vertex [see Eq.~(\ref{quartic})] and either a $(r^\dagger r)^2$
or $\he^\dagger\he r^\dagger r$ vertex.
We will ignore these diagrams in what follows because 
they are probably suppressed by one logarithmic factor
compared  with the 4-loop contributions 
discussed in Section 2. It should be mentioned, though, that
it is difficult to determine the largest number of  logarithmic factors
appearing in such 4-loop integrals, especially because some of these
are associated with infrared divergences for $M_r\!\to\! 0$.

Interestingly, the interactions of the $\Phi_8$ lead to an electron mass induced at 4
loops, as shown in Figure~\ref{fig:electron-mass}. Recall that the $r$ interactions 
by themselves allowed an electron mass only at 5 loops. It is thus likely that the
4-loop diagrams of Figure~\ref{fig:electron-mass} represent the dominant contributions
to the electron mass. Hence, the 
$\lambda_{11}\lambda'_{11}$ product may be smaller than the value derived from 
Eq.~(\ref{electron}) without affecting the 
electron mass, which somewhat relaxes the limits on the $r$ leptoquark (see Section 5).
On the other hand, these 4-loop diagrams include fewer than four  
logarithmic factors, while the 5-loop contributions are enhanced by the large 
number of diagrams, so that without a detailed computation it cannot be ruled out 
that the two contributions are comparable for sizable ranges of parameters.
Assuming that the diagrams in Figure~\ref{fig:electron-mass} dominate and lead to three
logarithmic factors ({\it i.e.}, one less than a 4-loop rainbow diagram, as suggested by 
the 2-loop computations presented in the Appendix), we find
\be
m_e \approx \lambda_{11}\, \lp_{11} \, 
\lp_{12}\,\lp_{22}\,\lp_{23}\,\lp_{33} \, \kappa_1^* \, c^{\prime\, *} 
\frac{y_\phi\langle \phi \rangle}{M_\Psi} 
\, v_\inh\,\epsilon^{(4)}_\Phi  ~,
\label{eq:electron}
\ee
where the 4-loop factor is 
\be
\epsilon^{(4)}_\Phi \sim \frac{N_c^2}{(16\pi^2)^4}\ln^3\! \left(\frac{M_\Psi^2}{M_r^2}\right) ~.
\label{four-loop-factor}
\ee 
The $m_e/v_\inh$ ratio is correctly reproduced for
\be
\lambda_{11}\, \lp_{11} \, \lp_{12}\,\lp_{22}\,\lp_{23}\,\lp_{33} \, 
\, \kappa_1^* \, c^{\prime\, *} \,  \sim 2 ~.
\label{electron-constraint}
\ee


\subsection{Strange and down quark masses}

The strange and down quarks remain massless until the 
$U(2)_d$ symmetry is broken. 
One possibility for breaking that chiral symmetry is to introduce 
some vectorlike fermions $\Upsilon^k$ which transform 
as $(1,2,+3/2)$ under $SU(3)_c\times SU(2)_W\times U(1)_Y$.
It is sufficient to include two such fermions: $k =1,2$.
Their most general Yukawa interactions with the fields introduced here are
\be
\eta_{jk} \, r\, \overline{d}_R^{\, j} \Upsilon^k   + \mathrm{H.c.}
\label{eq:upsilon}
\ee
An $U(2)_d$ transformation allows us to take $\eta_{11} = 0$.  We may also redefine the phases 
of the $s_R$ and $d_R$ fields such that 3 combinations of $\eta_{ij}$ parameters are real, 
but it is more convenient to do so after we compute the strange and down quark masses.

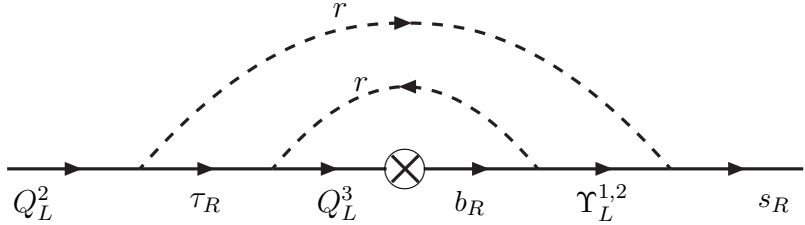
\begin{figure}[t]
\vspace*{-0.5cm}
\begin{center} 
{
\unitlength=1 pt
\SetScale{1}\SetWidth{1.2}      
\normalsize    
{} \allowbreak
\begin{picture}(300,90)(0,-10)
\ArrowLine(0,0)(50,0)\ArrowLine(50,0)(100,0)\ArrowLine(100,0)(143,0)\ArrowLine(157,0)(200,0)
\ArrowLine(200,0)(250,0)\ArrowLine(250,0)(300,0)
\DashCurve{(100,0)(135,28)(150,31)(165,28)(200,0)}{4}
\LongArrow(151,31)(149,31)
\DashCurve{(50,0)(120,51)(150,55)(180,51)(250,0)}{4}
\LongArrow(149,55)(151,55)
\Line(145,-5)(155,5)\Line(155,-5)(145,5)
\put(150.5,0){\circle{15}}
\Text(10,-13)[c]{$Q_L^2$}\Text(290,-13)[c]{\small $s_R$}
\Text(75,-13)[c]{$\tau_R$}\Text(225,-13)[c]{\small $\Upsilon_L^{1,2}$}
\Text(125,-13)[c]{$Q_L^3$}\Text(175,-13)[c]{\small $b_R$}
\Text(134,32)[c]{$r$}\Text(126,60)[c]{$r$}
\end{picture}
}
\end{center}
\vspace*{-0.2cm}
\caption{Strange mass induced by a 3-loop ``rainbow'' diagram. The $\otimes$ symbol represents 
the $b$ mass induced at 1-loop as in Fig.~\ref{fig:bmass}.
}
\vspace*{0.6cm}
\label{fig:strange}
\end{figure}

The strange-quark mass is generated at 3 loops by the diagram shown in Figure~\ref{fig:strange}.  
The contributions to the down-type quark mass matrix from this type of 3-loop rainbow diagrams
are approximately given by
\be
M_d[r r\Phi_8] \approx
\begin{pmatrix}
0 & 0 & 0 \\
0 & \left(\eta_{21}\,\eta_{31}^* + \eta_{22}\,\eta_{32}^*\right) \lp_{23} 
& \left(\eta_{21}\,\eta_{31}^* + \eta_{22}\,\eta_{32}^*\right)  \lp_{33} \\[0.5em]
0 & \left(\left|\eta_{31}\right|^2 + \left|\eta_{32}\right|^2\right) \lp_{23} 
&  \left(\left|\eta_{31}\right|^2 + \left|\eta_{32}\right|^2\right) \lp_{33} 
\end{pmatrix} \lp_{33} \, m_b \, \epsilon_r^{(2)}  ~,
\label{eq:strange}
\ee
where $\epsilon_r^{(2)}$ is the dimensionless 2-loop integral for a rainbow diagram
[see Eq.(\ref{epsilon-r2})]. We have assumed here that the vectorlike fermions $\Upsilon^{1,2}$
have negligible masses compared to $M_r$.
Since there is a 1-loop contribution to the $b$-quark mass (the 33 element of $M_d$), 
the $s$-quark mass is approximately given by the 22 element of $M_d[r r\Phi_8]$.
An $s_R$ field redefinition allows us to make the $\eta_{21}\,\eta_{31}^* + \eta_{22}\,\eta_{32}^*$
combination real and positive, so that all entries in Eq.~(\ref{eq:strange}) are positive.

The $m_s/m_b \sim 1/50$ mass ratio requires a product of dimensionless couplings 
to be larger than unity:
\be
\left(\eta_{21}\,\eta_{31}^* + \eta_{22}\,\eta_{32}^*\right) \lp_{23} \lp_{33}
\sim 8 ~.
\ee
However, given that several couplings are
involved here, none of them needs to be substantially larger than unity, and therefore we 
do not need to worry about departures from  perturbativity.

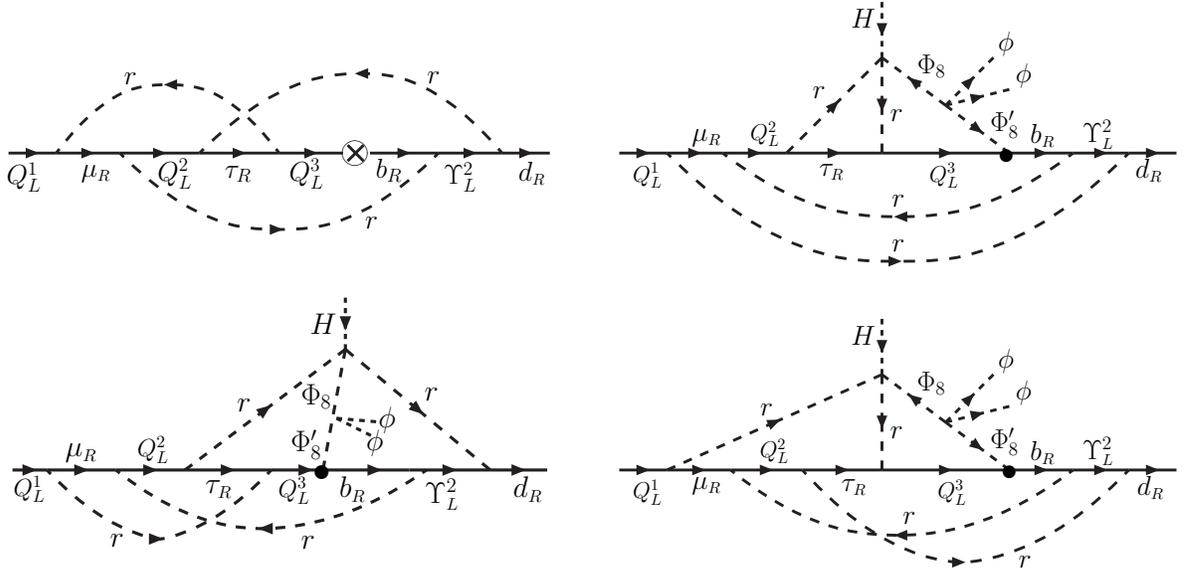
\begin{figure}[t]
\begin{center} 
\unitlength=1 pt
\SetScale{1}\SetWidth{1.8
}      
\normalsize\Large
{} \allowbreak
\begin{picture}(500,213)(48,-30)
\put(40,120){\scalebox{0.6}{
\ArrowLine(30,0)(60,0)\ArrowLine(60,0)(100,0)\ArrowLine(100,0)(150,0)\ArrowLine(150,0)(200,0)
\ArrowLine(200,0)(240,0)\ArrowLine(257,0)(300,0)\ArrowLine(300,0)(340,0)\ArrowLine(340,0)(370,0)
\DashCurve{(150,0)(220,46)(250,50)(280,46)(340,0)}{7}
\LongArrow(251,50)(249,50)
\DashCurve{(100,0)(170,-45)(200,-48)(230,-45)(300,0)}{7}
\LongArrow(199,-48)(201,-48)
\DashCurve{(60,0)(100,37)(130,43)(160,37)(200,0)}{6.5}
\LongArrow(131,43)(129,43)
\Line(243,-5)(253,5)\Line(253,-5)(243,5)
\put(248.5,0){\circle{16}}
\Text(40,-15)[c]{$Q_L^1$}\Text(85,-13)[c]{$\mu_\inr$}
\Text(135,-13)[c]{$Q_L^2$}\Text(175,-13)[c]{$\tau_R$}
\Text(218,-13)[c]{$Q_L^3$}\Text(270,-11)[c]{$b_R$}
\Text(315,-15)[c]{$\Upsilon^2_L$}\Text(360,-13)[c]{$d_\inr$}
\Text(259,-44)[c]{$r$}\Text(107,47)[c]{$r$}\Text(298,48)[c]{$r$}
}}
\put(280,120){\scalebox{0.6}{
\ArrowLine(15,0)(45,0)\ArrowLine(45,0)(80,0)\ArrowLine(80,0)(120,0)
\ArrowLine(120,0)(180,0)\ArrowLine(180,0)(260,0)\ArrowLine(260,0)(300,0)
\ArrowLine(300,0)(335,0)\ArrowLine(335,0)(365,0)
\DashArrowLine(180,60)(180,0){5.5}\DashArrowLine(180,95)(180,60){4}
\DashArrowLine(120,0)(180,60){5.5}
\DashArrowLine(220,30)(260,0){5.5}\DashArrowLine(220,30)(180,60){5.5}
\DashArrowLine(220,30)(260,40){4}\DashArrowLine(220,30)(250,60){4}
\Text(33,-14)[c]{\large $Q_L^1$}\Text(70,10)[c]{$\mu_\inr$}\Text(107,12)[c]{\large $Q_L^2$}
\Text(149,-12)[c]{$\tau_\inr$}
\Text(220,-14)[c]{\large $Q_L^3$}\Text(282,11)[c]{$b_\inr$}\Text(317,12)[c]{$\Upsilon_\inl^2$}
\Text(349,-11)[c]{$d_\inr$}
\Text(190,24)[c]{$r$}\Text(169,85)[c]{$H$}  
\Text(141,38)[c]{$r$}\Text(212,55)[c]{$\Phi_8$}\Text(258,18)[c]{$\Phi_8^\prime$}
\Text(259,-2)[c]{\LARGE $\bullet$}
\Text(270,49)[c]{$\phi$}\Text(259,69)[c]{$\phi$}
\DashCurve{(80,0)(150,-36)(190,-40)(230,-36)(300,0)}{7}
\LongArrow(191,-40)(189,-40)
\DashCurve{(45,0)(150,-65)(190,-68)(230,-65)(335,0)}{7}
\LongArrow(189,-68)(191,-68)
\Text(191,-58)[c]{$r$}\Text(191,-30)[c]{$r$}
}}
\put(40,0){\scalebox{0.65}{
\ArrowLine(30,0)(50,0)\ArrowLine(50,0)(90,0)\ArrowLine(90,0)(130,0)\ArrowLine(130,0)(180,0)
\ArrowLine(180,0)(220,0)\ArrowLine(220,0)(260,0)\ArrowLine(260,0)(307,0)\ArrowLine(307,0)(340,0)
\DashLine(223,70)(210,0){5.5}\DashArrowLine(223,100)(223,70){3}
\DashArrowLine(130,0)(223,70){5.5}
\DashArrowLine(223,70)(307,0){5.5}
\put(46,0){\DashLine(172,30)(192,20){3}\DashLine(172,30)(195,28){3}
\Text(195,16)[c]{$\phi$}\Text(202,29)[c]{$\phi$} 
\Text(161,43)[c]{$\Phi_8$}\Text(155,15)[c]{$\Phi_8^\prime$}\Text(165,85)[c]{$H$}  
}
\Text(40,-11)[c]{\large $Q_L^1$}\Text(70,10)[c]{$\mu_\inr$}\Text(112,12)[c]{\large $Q_L^2$}
\Text(150,-11)[c]{$\tau_\inr$}\Text(194,-12)[c]{\large $Q_L^3$}\Text(228,-11)[c]{$b_\inr$}
\Text(280,-14)[c]{$\Upsilon_\inl^2$}
\Text(330,-11)[c]{$d_\inr$}
\Text(274,44)[c]{$r$}
\Text(165,38)[c]{$r$}
\Text(210,-2)[c]{\LARGE $\bullet$}
\DashCurve{(50,0)(100,-36)(115,-40)(130,-36)(180,0)}{6}
\LongArrow(114,-40)(116,-40)
\DashCurve{(90,0)(160,-33)(175,-34)(190,-33)(270,0)}{5.5}
\LongArrow(176,-34)(174,-34)
\Text(91,-41)[c]{$r$}\Text(202,-42)[c]{$r$}
}}
\put(280,0){\scalebox{0.6}{
\ArrowLine(15,0)(45,0)\ArrowLine(45,0)(85,0)\ArrowLine(85,0)(130,0)
\ArrowLine(130,0)(180,0)\ArrowLine(180,0)(260,0)\ArrowLine(260,0)(300,0)
\ArrowLine(300,0)(335,0)\ArrowLine(335,0)(365,0)
\DashArrowLine(180,60)(180,0){5.5}\DashArrowLine(180,95)(180,60){4}
\DashArrowLine(45,0)(180,60){5.5}
\DashArrowLine(220,30)(260,0){5.5}\DashArrowLine(220,30)(180,60){5.5}
\DashArrowLine(220,30)(260,40){4}\DashArrowLine(220,30)(250,60){4}
\Text(33,-14)[c]{\large $Q_L^1$}\Text(70,-10)[c]{$\mu_\inr$}\Text(113,12)[c]{\large $Q_L^2$}
\Text(163,-11)[c]{$\tau_\inr$}\Text(282,11)[c]{$b_\inr$}\Text(317,12)[c]{$\Upsilon_\inl^2$}
\Text(225,-14)[c]{\large $Q_L^3$}\Text(350,-11)[c]{$d_\inr$}
\Text(190,24)[c]{$r$}\Text(169,85)[c]{$H$}  
\Text(108,38)[c]{$r$}\Text(212,55)[c]{$\Phi_8$}\Text(258,18)[c]{$\Phi_8^\prime$}
\Text(261,-2)[c]{\LARGE $\bullet$}
\Text(270,49)[c]{$\phi$}\Text(259,69)[c]{$\phi$}
\DashCurve{(85,0)(150,-36)(190,-41)(230,-36)(300,0)}{6}
\LongArrow(191,-41)(189,-41)
\DashCurve{(130,0)(210,-55)(230,-58)(250,-55)(335,0)}{7}
\LongArrow(229,-58)(231,-58)
\Text(271,-58)[c]{$r$}\Text(198,-30)[c]{$r$}
}}
\end{picture}
\end{center}
\caption{Down-quark mass induced at  4 loops. In the first diagram, the $\otimes$ represents the  
1-loop $b$-mass insertion of Fig.~\ref{fig:bmass}, and the upper two $r$ lines do not intersect. 
The last two diagrams are also nonplanar (the lower two $r$ lines do not intersect).  As before the $\bullet$ indicates a vertex obtained by integrating out the $\Psi$ fermion. 
}
\label{fig:downmass}
\end{figure}

The down-quark mass is generated at 4 loops through the diagrams 
shown in Figure~\ref{fig:downmass}, and is given by 
\be
m_d \approx 
\lp_{12}\,\lp_{22}\,\lp_{23}\,\lp_{33} \, \eta_{32}^* \, \eta_{12}
\left[ m_b\, \epsilon_\Upsilon^{(3)} + y_\phi \, v_\inh \kappa^\prime c\, c^\prime 
\left(\frac{\langle \phi \rangle}{M_\Psi} \right)^{\! 3} 
\epsilon^{(4)}_\Upsilon \right] ~.
\label{eq:down}
\ee
The first term in the paranthesis represents the first diagram in 
Figure~\ref{fig:downmass}, and involves a 3-loop nonplanar integral 
$\epsilon_\Upsilon^{(3)} $.
The second term represents the sum of the last three diagrams, and involves
a 4-loop dimensionless integral $\epsilon^{(4)}_\Upsilon$. 
Although we have not calculated $\epsilon^{(3)}_\Upsilon$
and  $\epsilon^{(4)}_\Upsilon$, the second term is likely to dominate
because it is enhanced by a factor of $N_c^2$: 
$\epsilon^{(4)}_\Upsilon \sim \epsilon^{(4)}_\Phi$, with $\epsilon^{(4)}_\Phi$
given in Eq.~(\ref{four-loop-factor}).
Similar 4-loop diagrams contribute to all $1i$ and $i1$ ($i=1,2,3$)
entries of the down-type mass matrix.

Using a phase redefinition of $d_R$, we take $m_d > 0$ in
Eq.~(\ref{eq:down}). This implies $\eta^*_{32}\eta_{12} c^\prime > 0$,
assuming that the first term is negligible. 
The interactions (\ref{eq:upsilon}) and the phase redefinitons discussed in this section finally break
the chiral symmetry $[U(3)]^5$ of the standard model fermions down to  
$U(1)_L \times U(1)_Q$, corresponding to lepton and quark number respectively.  

The $m_d/v_\inh \approx 1.4 \times 10^{-5}$ ratio also requires a product of dimensionless couplings to 
be large:
\be
\lp_{12}\,\lp_{22}\,\lp_{23}\,\lp_{33} \, \eta_{32}^* \, \eta_{12} \,
y_\phi \, \kappa^\prime c\, c^\prime \sim O(10)
\ee
for $\langle \phi \rangle \approx M_\Psi$. As in the case of $m_s$, the large number of couplings 
allows the product to be large even if no coupling is substantially larger than unity.

\begin{table}[t]
\renewcommand{\arraystretch}{1.6}
\centerline{
\label{tab:charges}
\begin{tabular}{|c||cccc|ccc|}
\hline
& $H$ & $\phi$ & $\Psi_{L,R}$ & $r$ & $\Phi_8$ & $\Phi_8^\prime$ & 
$\Upsilon^{1,2}_{L,R}$\\
\hline       \hline       
$SU(3)_c$           & $1$ & $1$ & $3$  & $3$   & $8$   & $8$    & $1$  \\
$SU(2)_W$           & $2$ & $1$ & $2$ & $2$   & $2$   & $2$    & $2$          \\
$U(1)_Y$          & \ $+1/2$ \ & \ $0$ \ & \ $+1/6$ \ & \ $+7/6$ \ & \ $+1/2$ \ & \ $-1/2$ \ & \ $+3/2$ \ \\
\hline
global $U(1)_H$ & $+1$ & $-1$& $-1$   & $0$  & $+1$   & $+1$    & $0$  \\
\hline
spin & 0 & 0 & 1/2 & 0 & 0 & 0 & 1/2 \\
\hline
\end{tabular}
}
\caption{Charges of scalars and vectorlike quarks. $H$ breaks the electroweak symmetry, 
$\phi$ and $\Psi$ communicate the breaking to the $t$ quark, $r$ communicates it to the 
charged leptons and the $c$ and $u$ quarks, while the fields on the right-hand side are 
responsible for down-type quark masses. 
}
\end{table}
%

\begin{figure}[h!]
\begin{center} 
\scalebox{.6}{
\unitlength=1 pt
\SetScale{1}\SetWidth{1.1}      
\normalsize\Large    
{} \allowbreak
\begin{picture}(400,385)(-130,-55)
\put(97,300){\circle{30}}\put(255,225){\circle{30}}
\put(97,145){\circle{30}}\put(255,75){\circle{30}}
\put(97,-5){\circle{30}}\put(255,-5){\circle{30}}
\ArrowLine(115,290)(235,230)\ArrowLine(118,297)(238,237)
\ArrowLine(235,220)(115,152)\ArrowLine(238,213)(118,145)
\ArrowLine(118,141)(236,83)\ArrowLine(117,133)(236,75)
\ArrowLine(235,70)(113,4)\ArrowLine(239,64)(117,-2)
\ArrowLine(115,-5)(235,-5)\ArrowLine(115,-11)(235,-11)
\Text(97,300)[c]{$t$}\Text(255,225)[c]{$\tau$}
\Text(97,145)[c]{$c$}\Text(255,75)[c]{$\mu$}
\Text(97,-5)[c]{$u$}\Text(255,-5)[c]{$e$}
\put(-160,0){
\Text(39,-60)[c]{\parbox{7cm}{\# of loops}}
\Text(-35,-5)[c]{4}
\Text(-35,75)[c]{3}\Text(-35,145)[c]{2}
\Text(-35,225)[c]{1}\Text(-35,300)[c]{0}
\multiput(-7,310)(.85,0){2}{\vector(0,-1){350}}
\LongArrow(-6.5,-35)(-6.5,-40)
}
\put(-75,225){\circle{30}}
\put(-75,75){\circle{30}}
\put(-75,-5){\circle{30}}
\ArrowLine(75,298)(-58,240)\ArrowLine(75,289)(-55,232)
\qbezier(230,225)(90,202)(-60,90)\ArrowLine(81,176)(78,175)
\qbezier(230,226)(90,203)(-60,91)
\DashArrowLine(95,280)(95,165){10}\ArrowLine(231,72)(-56,5)
\DashCurve{(77,285)(65,260)(55,220)}{9}
\DashCurve{(55,220)(50,185)(47,150)}{9}
\DashCurve{(47,150)(60,60)(77,9)}{9}\ArrowLine(49.5,133)(50,130.5)
\Text(-75,225)[c]{$b$}\Text(-75,75)[c]{$s$}\Text(-75,-5)[c]{$d$}
\ArrowLine(-74,204)(-74,97)\Text(-44,155)[c]{$\Upsilon,r,r$}
\qbezier(-88,204)(-140,102)(-88,10)
\qbezier(-87,204)(-139,101)(-87,9)
\Text(-124,30)[c]{$\Upsilon,r,r$}
\ArrowLine(-113,103)(-113,101)
\Text(185,278)[c]{$r$}\Text(170,250)[c]{$r$}
\Text(170,192)[c]{$r$}\Text(194,175)[c]{$r$}
\Text(185,120)[c]{$r$}\Text(171,93)[c]{$r$}
\Text(170,44)[c]{$r$}\Text(194,30)[c]{$r$}
\Text(185,4)[c]{$r$}\Text(185,-22)[c]{$r$}
\Text(15,246)[c]{$\Phi_8$}\Text(-5,280)[c]{$\Phi_8^\prime$}
\Text(110,226)[c]{$\Phi_8$}\Text(35,100)[c]{$\Phi_8$}
\Text(-15,135)[c]{$r$}\Text(-15,25)[c]{$r$}
\end{picture}
}
\end{center}
\caption{Loop-level of mass generation. Each line connecting a pair of fermions
indicates interactions that break their chiral symmetries. 
A fermion receives a mass provided at least two lines connect it to other
fermions which communicate with the Higgs sector
(the chiral symmetries of both its left- and 
right-handed components must be broken). 
}
\label{fig:map-complete}
\vspace*{.9cm}
\end{figure}
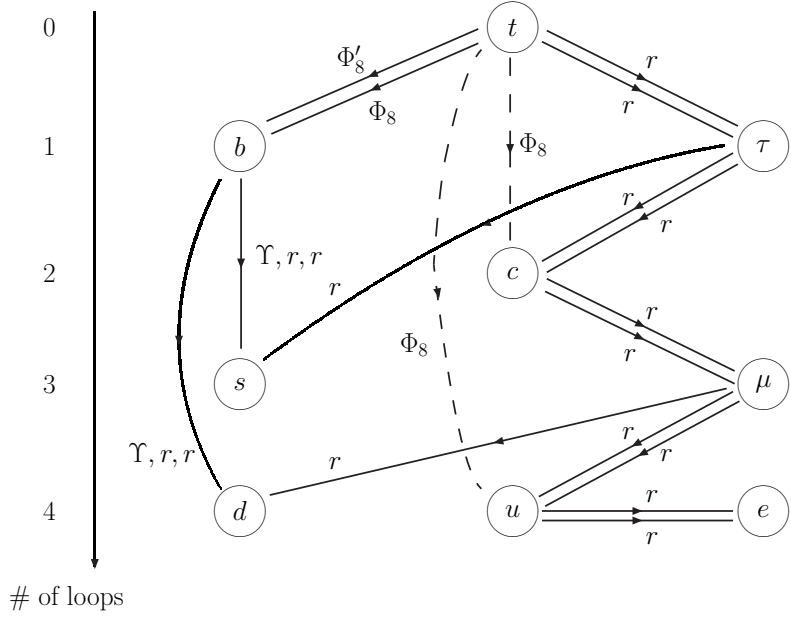

To summarize the field content of our model, the masses of the up-type quarks and charged leptons
are generated by the fields shown on the left-hand side of Table~1, 
while the masses of the down-type quarks also require the fields shown on the 
right-hand side of Table~1.
The mechanism of fermion mass generation is schematically depicted in 
Figure~\ref{fig:map-complete}.

\subsection{CKM matrix}
\label{sec:ckm}

In the previous sections we have determined the leading piece of each entry in 
the mass matrix for both the up- and down-type quarks.  The transformations necessary to go to the mass eigenstate basis will determine the CKM matrix.  The mass terms for the quarks are $\overline{q}_R M_q q_L$ ($q=u,d$).  Working in the regime where the charm quark gets its mass predominantly from diagrams involving $\Phi_8$, so that (\ref{eq:charmtwo}) dominates over (\ref{eq:charmone}), we find
\be
M_u\approx \begin{pmatrix}
m_u & \dfrac{\kappa_1}{\kappa_2}\, m_c & \dfrac{\kappa_1\lp_{33}}{\kappa_2\lp_{23}}\, m_c \\[3ex]
a_u m_u & m_c & \dfrac{\lp_{33}}{\lp_{23}}\, m_c \\[3ex]
\dfrac{\myl_{33}}{\myl_{23}}\left( a_u - \dfrac{\lambda_{22}}{\lambda_{12}} \right)
 m_u \;\; & \;\;\;\; \dfrac{\kappa_3}{\kappa_2}\, m_c \;\; \;\; & m_t
\end{pmatrix}~~,
\ee
where we defined
\be
a_u \equiv \dfrac{\myl_{22}^2+\myl_{23}^2+\myl_{33}^2}{\myl_{12}\myl_{22}} > 0 ~.
\ee
The only complex entries in $M_u$ are those involving $\kappa_1$ or $\kappa_2$.
The down-type quark mass matrix is given by
\be
M_d\approx \begin{pmatrix}
m_d & \; \dfrac{\lambda^{\prime\,2}_{22}+\lambda^{\prime\,2}_{23}+\lambda^{\prime\,2}_{33}}{\lp_{12}\lp_{22}}\, m_d \;\;\;  & \dfrac{\lp_{33}(\lambda^{\prime\,2}_{23}+\lambda^{\prime\,2}_{33})}{\lp_{12}\lp_{22}\lp_{23}}\, m_d \\[4ex]
 a_d\, m_d & m_s & \dfrac{\lp_{33}}{\lp_{23}}\, m_s \\[2ex]
a_d\, a_s\, m_d 
&  a_s\, m_s & m_b
\end{pmatrix}~,
\ee
where, for convenience we introduced the notation
\bear
&& a_d = \dfrac{\eta_{21}\, \eta_{31}^*+\eta_{22}\, \eta_{32}^*}{\eta_{12}\, \eta_{32}^*} ~,
\nonumber \\ [0.6em]
&& a_s =  \dfrac{\left|\eta_{31}\right|^2+\left|\eta_{32}\right|^2}
{\eta_{21}\, \eta_{31}^*+\eta_{22}\,\,\eta_{32}^*} > 0 ~.
\eear
Note that $a_d$ is the only complex parameter, and the only complex entries in $M_d$ are the 21 and 31 ones.

The CKM matrix is determined by the unitary transformations, 
$V_{u_L}$ and $V_{d_L}$, of the left-handed quark fields that 
diagonalize the mass matrices (more precisely, the $3\times 3$ 
matrices $V_{q_L} M_q^\dagger M_q V_{q_L}^\dagger$ for $q=u,d$ are diagonal).  
For the up-type quarks this transformation is
\be
V_{u_L}\approx\begin{pmatrix}
1 & \; - \dfrac{\kappa_1/\kappa_2 + a_u }{1 + |\kappa_1/\kappa_2|^2} \, \dfrac{m_u}{m_c} \;  & 0 \\[2.3ex]
\dfrac{\kappa_1^*/\kappa_2^* + a_u }{1 + |\kappa_1/\kappa_2|^2} \, \dfrac{m_u}{m_c} & 1 & -\dfrac{\kappa_3}{\kappa_2^*}\, \dfrac{m_c}{m_t} \\[2.3ex]
0 & \;\; \dfrac{\kappa_3}{\kappa_2}\, \dfrac{m_c}{m_t} \;\; & 1
\end{pmatrix} ~,
\ee
where we have ignored all quadratic corrections in $m_u/m_c$ or $m_c/m_t$ (these ratios are 2-loop factors, 
as shown in Figure~\ref{fig:map-complete}).
For the down-type quarks, we keep only linear terms in $m_s/m_b$ (which is a 2-loop factor) 
but we keep the quadratic terms in $m_d/m_s$ (which is a 1-loop factor):
\be
V_{d_L}\approx\begin{pmatrix}
1-\dfrac{\left|a_d\right|^2}{2}\left(\dfrac{m_d}{m_s}\right)^{\!\!2} & -a_d^* \dfrac{m_d}{m_s} 
& 0 
\\[2.6ex]
a_d\dfrac{m_d}{m_s} & \;\;\; 1-\dfrac{\left|a_d\right|^2}{2}\left(\dfrac{m_d}{m_s}\right)^{\!\!2} \;\;\; 
& -a_s\,\dfrac{m_s}{m_b}
\\[3.1ex]
a_d\,a_s\, \dfrac{m_d}{m_b} & a_s\, \dfrac{m_s}{m_b} & 1
\end{pmatrix}
\label{eq:rots}
\ee

The CKM matrix is then given by
\be
V_{u_L} V_{d_L}^\dagger \approx \begin{pmatrix}
1-\dfrac{|a_d|^2}{2}\left(\dfrac{m_d}{m_s}\right)^{\!\!2} & a_d^*\dfrac{m_d}{m_s} & a_d^{*}\,a_s\dfrac{m_d}{m_b}  \\[2.6ex]
-a_d\dfrac{m_d}{m_s} & \;\;\; 1-\dfrac{|a_d|^2}{2}\left(\dfrac{m_d}{m_s}\right)^{\!\!2} \;\;\; 
& a_s\, \dfrac{m_s}{m_b}-\dfrac{\kappa_3}{\kappa_2^*}\, \dfrac{m_c}{m_t} \\[3.1ex]
 - a_d\dfrac{\kappa_3}{\kappa_2}\dfrac{m_c m_d}{m_t m_s}  & -a_s\, \dfrac{m_s}{m_b}  
+ \dfrac{\kappa_3}{\kappa_2}\, \dfrac{m_c}{m_t}   & 1
\\
\end{pmatrix} ~.
\ee
All off-diagonal entries of this matrix are  complex, but one may absorb four phases in 
the $u_L^i$ and $d_L^i$ fields, leaving complex phases only into the 13 and 31 entries.  
The result is the CKM matrix in the Wolfenstein parametrization \cite{Wolfenstein:1983yz},
\be
V_{\rm CKM} \approx \begin{pmatrix}
1 - \dfrac{\lambda^2}{2}  & \lambda  & \; A \lambda^3 \left( \rho - i \eta \right) \\[3ex]
- \lambda &  \; 1 - \dfrac{\lambda^2}{2} \; & A \lambda^2 \\[3ex]
\; A \lambda^3 \left( 1 - \rho - i \eta \right) \;\;\; & - A \lambda^2 & 1 \\
\end{pmatrix} ~,
\ee
with the Wolfenstein parameters given in terms of our combinations of couplings by
\bear
&& \lambda = |a_d| \frac{m_d}{m_s}  ~,
\nonumber \\[2ex]
&& A = \frac{1}{\lambda^2} 
\left| a_s \frac{m_s}{m_b} - \frac{\kappa_3}{\kappa_2^*} \frac{m_c}{m_t}\right|  ~,
\nonumber \\[2ex]
&& \rho = \frac{a_s}{A^2\lambda^4} \frac{m_s}{m_b} 
\left[ a_s \frac{m_s}{m_b} - {\rm Re}\left(\frac{\kappa_3}{\kappa_2}\right) \frac{m_c}{m_t}\right]  ~,
\nonumber \\[2ex]
&& \eta =  \frac{a_s}{A^2\lambda^4} 
{\rm Im}\left(\frac{\kappa_3}{\kappa_2}\right) \frac{m_s}{m_b}  \frac{m_c}{m_t}  ~,
\eear
These equations can be inverted:
\bear
&& |a_d| =  \lambda \, \frac{m_s}{m_d} \, \approx \, 4.4  ~,
\nonumber \\[2ex]
&& a_s = A \lambda^2 \sqrt{\rho^2 + \eta^2} \,\frac{m_b}{m_s} \, \approx \, 0.84~,
\nonumber \\[2ex]
&& \frac{\kappa_3}{\kappa_2} = 
\frac{A \lambda^2}{\sqrt{\rho^2 + \eta^2}} \, \frac{m_t}{m_c} 
\left[ \eta^2 + \rho^2 - \rho + i \eta \right] 
\, \approx \, -1.6 + 8.9 \, i ~,
\eear
where the numerical values used here are
$\lambda \approx 0.227$, $A \approx 0.818$, $\rho \approx 0.22$, and $\eta \approx 0.34$ \cite{Yao:2006px}.
The ratios $|a_d|$ and Im$(\kappa_3/\kappa_2)$ are larger than order one,
but overall the CKM matrix elements are well reproduced in our model for reasonable values 
of parameters.

\section{Experimental Constraints}
\label{sec:constraints}
\setcounter{equation}{0}

The domino mechanism described above works at any scale provided there is some 
separation between the masses of the domino particles and the cutoff scale 
(set by the mass of the $\Psi$ fermion).  However, if the mass scales are 
low enough it may be possible to directly produce some of the new states 
or to probe them indirectly by their effects on rare processes.  For instance, 
leptoquarks induce rare meson decays, rare $\mu$ and $\tau$ decays, $\mu\to e$ 
conversion in nuclei, meson anti-meson mixing and other processes~\cite{Davidson:1993qk}.  The constraints from rare 
processes typically bound the ratios $\lambda_{ij}/M_r$ or $\lambda_{ij}^2/M_r$.  
Since we know the approximate size of the couplings necessary to give the 
correct quark and lepton masses, we derive a lower bound on the leptoquark
mass.  

The conversion of $\mu\to e$ in nuclei can take place at tree level through exchange 
of the leptoquark. At low-energy, the relevant piece of the effective Lagrangian,
after a Fierz transformation, is 
\bear
\frac{1}{4M_r^2} \!\!\!\! && \!\!\!\!
\left[ \bar{u} u \left( \lambda_{11}^\prime \lambda_{12} \, \bar{e}_R \mu_L  
+ \lambda_{11} \lambda_{12}^\prime \, \bar{e}_L \mu_R \right) 
- \bar{u}\gamma_\mu u  \left( \lambda_{11} \lambda_{12} \, \bar{e}_L \gamma^\mu \mu_L  
+ \lambda_{11}^\prime \lambda_{12}^\prime \, \bar{e}_R \gamma^\mu \mu_R\right) 
\rule{0mm}{3.9mm} \right.
\nonumber \\ [1mm]
&& \left. - \; 
\lambda_{11}^\prime \lambda_{12}^\prime \, \bar{d} \gamma_\mu d \;
\bar{e}_R \gamma^\mu \mu_R\rule{0mm}{3.9mm} \right] ~.
\eear 
The above terms involving $u$ quarks arise from the exchange of the $r$
component carrying  weak isospin +1/2, while the terms involving $d$ quarks 
arises from the one carrying weak isospin $-1/2$.
Following Ref.~\cite{Kitano:2002mt}, we find that the above four-fermion 
interactions give the following rate for coherent $\mu \to e$ conversion
in nuclei:
\be
\Gamma(\mu \to e) = \frac{m_\mu^5}{4 M_r^4} 
\left\{ \myl_{11}^2 \left[\lp_{12} S_0 - \myl_{12} \left(2V^{(p)}+V^{(n)}\right) \right]^2 + 
\lambda_{11}^{\prime \, 2} \left[ \myl_{12} S_0 
- 3 \lp_{12} \left(V^{(p)}+V^{(n)}\right) \right]^2 \rule{0mm}{3.9mm}\right\}
\ee
where $V^{(p)}$ $(V^{(n)})$ is the overlap integral of the proton (neutron) density 
and the electron and muon wavefunctions associated with vector operators, 
and $S_0$ is a combination of similar integrals for scalar operators. 
For Titanium, these are \cite{Kitano:2002mt}: $V^{(p)}=0.0396$, $V^{(n)}=0.0468$ 
and $S_0\approx 0.375$.
The experimental limit on muon conversion \cite{Yao:2006px} in Titanium is
\be
\frac{\Gamma(\mu\,\mathrm{Ti} \to e\,\mathrm{Ti})}{\Gamma(\mu\,\mathrm{Ti} \to\mathrm{capture})} 
< 4.3\times 10^{-12}.
\ee
Thus, we find a limit 
\be
M_r > 290\,\tev
\left[ \myl_{11}^2 \left(\myl_{12} - 3.0 \lp_{12} \right)^2 + 
\lambda_{11}^{\prime \, 2} \left( 3.0 \myl_{12}  
- 2.1 \lp_{12} \right)^2 \rule{0mm}{3.9mm} \right]^{1/4} ~.
\ee  
The Yukawa couplings here are in the mass eigenstate basis, whereas the ones 
introduced in section 2 are given in the weak eigenstate basis. Given that the two bases 
are roughly aligned, we will not make the distinction explicit in what follows.
The $m_u/m_\mu$ ratio requires $\myl_{12}\lp_{12}\sim (0.6)^2$,
as discussed in Section 2.
Likewise, the product $\myl_{11}\lp_{11}$ cannot be too small, or else the loop-generated 
electron mass will not be consistent with the measured value. Nevertheless,
several of the couplings in Eq.~(\ref{electron-constraint}) may be larger than unity, allowing
$\myl_{11}\lp_{11}\sim 0.1$.
For $\myl_{12}\approx \lp_{12} \approx 0.6$ and $\myl_{11}\approx \lp_{11}\approx 0.3$,
the mass limit is $M_r\!\!\gae\!\! 180$ TeV. A more judicious choice of couplings would relax
the mass limit: by tuning the couplings while keeping $\lambda_{11}\!\!\lae\! 2$, 
the limit becomes $M_r\!\!\gae\!\! 100$ TeV.

The scalar leptoquark $r$ has chirality violating couplings since it couples to both 
left- and right-handed quarks and leptons.  It may contribute, at tree level, 
to decays that are helicity suppressed in the standard model, such as the 
decays of the pseudoscalar mesons, but without the mass suppression.  
The new contribution to the decay amplitude interferes with the standard model  
amplitude, so the leptoquark contribution to the rate scales as $1/M_{r}^2$.  
The ratio of the helicity suppressed decay of the pion to the dominant mode  
is measured to be \cite{Yao:2006px}
\be
R \equiv \frac{\Gamma \left(\pi^+\to e^+ \nu\right)}{\Gamma\left(\pi^+\to \mu^+ \nu\right)} = \left( 1.230 \pm 0.004\right) \times 10^{-4},
\ee
and the SM prediction is \cite{Cirigliano:2007xi} 
$R_{SM}=\left(1.2352\pm 0.0001 \right)\times 10^{-4}$.  The contribution from exchange of an $r$ leptoquark is
\be
\frac{R_{LQ}}{R_{SM}} = 
\frac{1}{2\sqrt{2} G_F V_{ud}} \frac{m_\pi^2}{m_u+m_d} \frac{1}{M_r^2} \left(\frac{\myl_{11}\lp_{11}}{m_e} - \frac{\myl_{12}\lp_{12}}{m_\mu} \right).
\ee
Since the leptoquark enhances $R$, and the standard model prediction is 
already above the observed value, the constraint on the leptoquark is strong.  
At the  $95\%$ CL,
\be
\frac{M_r}{\sqrt{\myl_{11}\lp_{11}}}> 270\, \tev.
\ee
As discussed above, $\myl_{11}\lp_{11} \gae 0.1$ so that the pion decays require 
$M_r \gae 90$ TeV.

At 1-loop the leptoquark contributes to processes like $K-\overline{K}$ mixing or $\mu\to e\gamma$.  
Let us briefly discuss the former. The contribution to $K_L^0-K_S^0$ mass splitting, $\Delta m_K $, from box diagrams involving $r$ is
\be
\Delta m_K^{(LQ)} = \frac{\left(\lp_{12}\lp_{22}\right)^2}{M_r^2} \frac{f_K^2\, m_K}{192\pi^2} ~,
\ee
where $f_K\approx 159.8\,\Mev$ is the kaon decay constant, and the measured mass splitting is $\Delta m_K = (3.483 \pm 0.006)\times 10^{-12}\, \Mev$.  Since there are large long distance uncertainties in the calculation of the SM contribution to $\Delta m_K$ we will assume that the new physics contribution from $r$ boxes can be as large as $30\%$ of the measured value.  Using the values $\lp_{12}\approx 0.6$ and $\lp_{22}\approx 1.5$,
as suggested in section~\ref{sec:lepton}, this results in a bound of $M_r\!\! \gae\! 70$ TeV.  

Additional constraints on $r$ are set by lepton-flavor violating $K$ decays (such as $K^+\to \pi^+\,\mu^+\, e^- $),
rare $\tau$ decays, $D-\overline{D}$, $B_s-\overline{B_s}$ mixing, and other processes.  
However, the limit on $M_r$ coming from $\mu\to e$ conversion in nuclei is currently
the most stringent one. Thus,
an improvement in the experimental sensitivity on $\mu\to e$ conversion in nuclei 
may lead to the discovery of the $r$ leptoquark effects.

The constraints on the fields used to generate the $b$ quark mass, 
$\Phi_8$ and $\Phi_8^\prime$, are more model dependent.  Color-octet weak-doublet scalars
of this type have been discussed in Ref.~\cite{Manohar:2006ga, Gerbush:2007fe}. 
In our case the flavor structure of their interactions is different, 
predominately involving a 3rd generation left-handed quark and a right-handed 
quark of any generation.  Through the down-type quark mixing (\ref{eq:rots}),
$\Phi_8^\prime$ gives tree-level contributions to $K-\overline{K}$ mixing, 
and together $\Phi_8$ and $\Phi_8^\prime$ give loop contributions to 
$b\to s\gamma$.  Due to the number of small mixings that enter, the 
constraint from $K-\overline{K}$ mixing is very weak, $M_{8^\prime} \gae \mathcal{O}(10\,\gev)$.  
The $b\to s\gamma$ process involves fewer mixing insertions and has a 
stronger constraint.  The contribution of $\Phi_8$ is similar to that of a charged Higgs boson.   
However, because the $b$ quark mass itself is generated at 1 loop, 
$b\to s\gamma$ is not loop suppressed, but is suppressed by small model-dependent 
mixings.  Depending on these couplings,
the $\Phi_8$ mass may be below the TeV scale, 
making it accessible at the LHC.  
The color-octet scalar $\Phi_8$ would then be produced in pairs via its coupling to the gluon,
and the signal would be a pair of equal mass resonances, such as $(t\,\bar{t})\, (t\,\bar{t})$, 
$(t\,\bar{b})\, (b\,\bar{t})$,  $(j\,\bar{b})\, (b\,j)$, or $(b\,\bar{t})\, (j\,\bar{b})$,
where $j$ is a jet coming from an up or charm quark. Some of these signatures have been studied
in Ref.~\cite{Dobrescu:2007yp, Gerbush:2007fe}.  
Single $\Phi_8$ production is also possible via gluon fusion \cite{Gresham:2007ri}.
The $\Phi_8$ can also alter the decays of the top quark.

The vectorlike leptons $\Upsilon^{1,2}$ are harder to produce, but their decays 
(into a charged lepton and two jets via a virtual $r$ at tree level, 
or into $\tau\gamma$ at 1-loop) are easier to observe.


\section{Conclusions}

The repeated mass hierarchies amongst elementary fermions
is a long-standing mystery in particle physics. We have proposed
that the fermion masses are generated by loops involving 
other standard model fermions. 
Starting with only the top-quark being heavy at tree level, 
and introducing a single scalar (leptoquark) 
which couples the up-type quarks to 
the leptons, we have shown that all these fermions acquire mass 
in turn, each at a higher loop level than the previous one. 
The outcome of this \emph{domino mechanism} is that the 
$\tau$, $c$, $\mu$, $u$ and $e$ masses are generated at 
1, 2, 3, 4 and 5 loops, respectively. 
Unlike many other methods for generating the Yukawa couplings, 
we do not distinguish between the generations of standard 
fermions. Even the top quark need not be singled out by 
a symmetry: in the presence of a heavy vectorlike quark, 
the tree-level mass matrix of the up-type quarks has rank 
one, such that only the top gets a tree-level mass.

The mechanism may be extended to the down-type quarks 
by including some other `domino' particles. The model building 
aspects here involve more moving parts. We have described an explicit 
example where the bottom-quark mass is generated by a loop 
involving a pair of color-octet scalars and the top quark. 
A byproduct of these color octets is that the charm mass 
receives additional 2-loop contributions, and the electron mass 
is generated at 4 loops.
The strange- and down-quark masses arise through loops involving 
the scalar octets and a vectorlike lepton.
Altogether, this model induces  bottom and tau masses at 1 loop, 
a charm mass at 2 loops, muon and strange masses at 3 loops, 
and masses for the first generation at 4 loops. 
With all couplings of order unity, this generates the 
correct patterns of fermion masses and CKM matrix elements.

Our mechanism works equally well anywhere between the electroweak 
and Planck scales. There are however constraints on the masses of the 
new scalars from various flavor-changing processes. 
The leptoquark has to be heavier than about 100 TeV, and its effects may 
be discovered in future experiments searching for $\mu \to e$ conversion
in nuclei, or rare $K$ decays. The constraints on the color octets are far 
weaker, allowing for interesting signatures involving third-generation quarks
at the LHC.

Given the relatively high mass required for the leptoquark, 
our domino mechanism must be embedded in a larger theory that also 
addresses the stability of the electroweak scale. We expect that 
it is possible to construct a supersymmetric 
theory of this type\footnote{Related supersymmetric models can be found in 
Ref.~\cite{Babu:1989tv}.}.
Another possibility is that the Higgs doublet
is a bound state of the top quark with a vectorlike quark,
as in the top seesaw model \cite{Dobrescu:1997nm}. The discovery at 
colliders of superpartners or of particles involved in dynamical 
electroweak symmetry breaking could allow tests of the 
flavor effects induced by the domino particles.

In total there are 24 parameters of our model that are involved in generating 
the entries of the fermion mass matrices. Once the top mass is fixed, there 
are only predictions for 8 fermion masses and 4 CKM elements leaving many 
parameters free. It would be interesting to embed the domino mechanism into 
a grand unified theory, which would reduce sufficiently the number of parameters 
to allow definite comparisons with the experimental values.
Intriguingly, all scalars introduced in this paper 
fit into the $\overline{126}$ representation of $SO(10)$.

\bigskip
{\bf Acknowledgments:} We would like to thank Thomas Becher, Sekhar Chivukula, 
Andr\'e de Gouv\^ea, Ayres Freitas, Dave Kaplan, Andreas Kronfeld, Zoltan Ligeti, 
Michael Ramsey-Musolf, Chris Quigg, Scott Thomas and Koichi Yamawaki for useful comments. 
P.J.F. would like to thank the KITP for hospitality while part of this work was completed.  
This research was supported in part by the National Science Foundation under Grant No. PHY05-51164.
B.D. acknowledges the hospitality and support of the
Radcliffe Institute for Advanced Study during early stages of this work.
Fermilab is operated by Fermi Research Alliance, LLC, under Contract
DE-AC02-07CH11359 with the United States Department of Energy.

\renewcommand{\theequation}{A.\arabic{equation}}
\section*{Appendix: 2-loop integrals}
\setcounter{equation}{0}

In this Appendix we compute the 2-loop integrals that contribute
to the charm mass. Let us begin with the rainbow diagram 
of Fig.~\ref{fig:charm}:

\bear
\label{eq:rainbow}
\epsilon^{(2)}_r & = &  N_c \int\frac{d^4 k^\prime}{(2\pi)^4} \; 
\frac{M_\Psi^2}{k^{\prime 2} (k^{\prime 2} - M_r^2)} \;
\int\frac{d^4 k}{(2\pi)^4} \; 
\frac{1}{k^2 (k^2 - M_\Psi^2) \left[(k-k^\prime)^2 - M_r^2\right]}
\nonumber \\ [3mm] 
& = & \frac{N_c}{(16\pi^2)^2} \, \int_0^1 \! dx\, f\!\left(x,M_\Psi^2/M_r^2\right)
\eear
where we defined 
\bear
f(x,\,a)  & = & 
a\int^{\,1/(a\, x)}_{\,1/[a\, x(1-x)]} \!\!\! d t \,\, 
\frac{\ln t}{1-t}
\nonumber \\ [3mm] 
& = & a \left[\dilog \left(1-\frac{1}{a\,x}\right) -\dilog \left(1-\frac{1}{a\,x(1-x)}\right)\right]~.
\eear
For $M_\Psi^2/M_r^2 \gg 1$,
\be
\epsilon^{(2)}_r \approx \frac{N_c}{(16\pi^2)^2} \, \left[  \frac{1}{2} \ln^2\!\left( \frac{M_\Psi^2}{M_r^2} \right) + \ln\! \left(\frac{M_\Psi^2}{M_r^2}\right) + \frac{391}{400} \right]~.
\ee

\begin{figure}[t]\center
\psfrag{faffa}[t]{\hspace{1.5cm} \parbox[t]{3cm}{ \vspace{-0.1cm} $f$, $\widetilde{f}$}}
\psfrag{Massratio}[t]{\hspace{3cm} \parbox[t]{3cm}{ \vspace{-0.3cm}$M_\Psi/M_r$} }
\psfrag{0.00}{}
\psfrag{0.05}{\hspace{-.3cm} $0.05\; $\ }
\psfrag{0.10}{$0.1$}
\psfrag{0.15}{\hspace{-.3cm} $0.15\, $\ }
\psfrag{0.20}{$0.2$}
\psfrag{0.25}{\hspace{-.3cm} $0.25\,$\ }
\psfrag{0.30}{$0.3$}
\psfrag{20}{$20$}
\psfrag{40}{$40$}
\psfrag{60}{$60$}
\psfrag{80}{$80$}
\psfrag{100}{$100$}
\psfrag{ep2poverep1}{\hspace{-0.8cm} $\epsilon_\Phi^{(2)}/\left(\epsilon_r^{(1)}\right)^2$}
\psfrag{ep2roverep1}{\hspace{-0.8cm} $\epsilon_r^{(2)}/\left(\epsilon_r^{(1)}\right)^2$}
\psfrag{YYY}{\hspace{-1.5cm} $\dfrac{\text{2-loop}}{\left(\text{1-loop}\right)^2}$}
\psfig{file=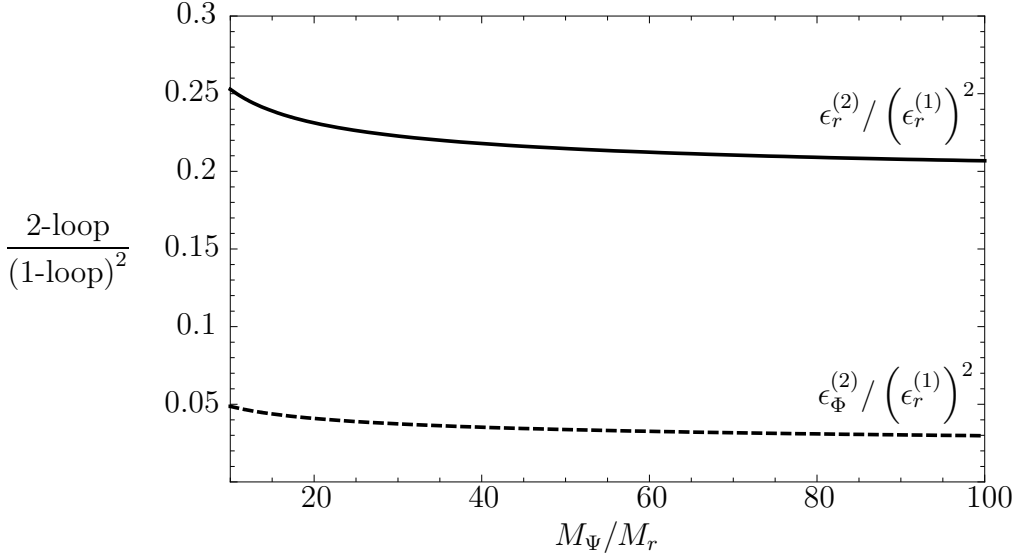,width=12.1cm,angle=0}
\caption{$\epsilon_r^{(2)}$ is the 2-loop factor generated by the rainbow diagram contribution to $m_c$ of Figure~\ref{fig:charm} and $\epsilon_\Phi^{(2)}$ is from the 2-loop diagram of Figure~\ref{fig:charm-mass}.  Here they are plotted relative to the 1-loop contribution to $m_\tau$ from Figure~\ref{fig:tau}.}
\label{fig:fofa}
\end{figure}

Let us now turn to the 2-loop diagram shown in Fig.~\ref{fig:charm-mass}. The 
associated 2-loop integral, which contributes to the charm mass 
as in Eq.~(\ref{eq:charmtwo}), is 
\bear
\label{eq:tent}
\epsilon^{(2)}_\Phi
& = & N_c \int\!\!\frac{d^4 k^\prime}{(2\pi)^4} \,
\frac{M_\Psi^2 \; \slash{\!\!\!k}^\prime}{k^{\prime 2} (k^{\prime 2} - M_8^2) 
(k^{\prime 2} - M_\Psi^2)} 
\int\!\!\frac{d^4 k}{(2\pi)^4}  \,
\frac{\slash{\!\!\!k}}{k^2 (k^2 - M_r^2)\left[(k-k^\prime)^2 - M_r^2\right]}
\nonumber \\ [3mm] 
& = & N_c \frac{M_\Psi^2}{16\pi^2 } \, \int^1_0 \! d x \int^1_0 \!\!\! d y \,\, 
\widetilde{I}_1 \left(M_\Psi,M_8,M_r\sqrt{(1/x-y)/(1-x)}\right) ~,
\eear
where $\widetilde{I}_1$ is the 1-loop integral given in Eq.~(\ref{eq:bloop}).
For $M_8 \ll M_r, M_\Psi$,
\be
\label{eq:tent2}
\epsilon^{(2)}_\Phi \approx \frac{N_c}{(16\pi^2)^2} \, \int_0^1 \! dx\, \left(1-x\right) f\!\left(x,M_\Psi^2/M_r^2\right)~.
\ee
For $M_\Psi^2/M_r^2 \gg 1$,
\be
\epsilon^{(2)}_\Phi \approx \frac{N_c}{(16\pi^2)^2} \left[\ln\!  \left(\frac{M_\Psi^2}{M_r^2}\right)  - \frac{\pi^2}{6}\right]~.
\ee
Both $\epsilon^{(2)}_r$ and $\epsilon^{(2)}_\Phi$ are shown in Figure~\ref{fig:fofa}.


\vfil

\bigskip\bigskip

\begin{thebibliography}{99} \frenchspacing

\bibitem{Yao:2006px}
  W.~M.~Yao {\it et al.}  [Particle Data Group],
  ``Review of particle physics,''
  J.\ Phys.\ G {\bf 33} (2006) 1, and 2007 partial update for the 2008 edition.

\bibitem{Babu:1989fg}
  For some brief reviews and lists of references, see: \\ 
   K.~S.~Babu and E.~Ma,
   ``Radiative Mechanisms For Generating Quark And Lepton Masses: Some Recent Developments'',
  Mod.\ Phys.\ Lett.\  A {\bf 4}, 1975 (1989); \\
  S.~M.~Barr,
  ``Radiative Fermion Mass Hierarchy in a Non-supersymmetric Unified Theory,''
  Phys.\ Rev.\  D {\bf 76}, 105024 (2007)
  [arXiv:0706.1490 [hep-ph]]; \\
  L.~E.~Ibanez,
  ``Radiative Fermion Masses In Grand Unified Theories,''
  Nucl.\ Phys.\  B {\bf 193}, 317 (1981); \\
  B.~A.~Dobrescu,
  ``Fermion masses without Higgs: A Supersymmetric technicolor model,''
  Nucl.\ Phys.\  B {\bf 449}, 462 (1995)
  [arXiv:hep-ph/9504399]; \\
  T.~Appelquist, Y.~Bai and M.~Piai,
  ``Quark mass ratios and mixing angles from SU(3) family gauge symmetry,''
  Phys.\ Lett.\  B {\bf 637}, 245 (2006)
  [arXiv:hep-ph/0603104].

\bibitem{Georgi:1972hy}
  H.~Georgi and S.~L.~Glashow,
  ``Attempts to calculate the electron mass,''
  Phys.\ Rev.\  D {\bf 7}, 2457 (1973); 
  ``Spontaneously broken gauge symmetry and elementary particle masses,''
  Phys.\ Rev.\  D {\bf 6}, 2977 (1972). \\
  S.~Weinberg,
  ``Electromagnetic and weak masses,''
  Phys.\ Rev.\ Lett.\  {\bf 29}, 388 (1972). 

\bibitem{Balakrishna:1988ks}
  B.~S.~Balakrishna,
  ``Fermion Mass Hierarchy From Radiative Corrections,''
  Phys.\ Rev.\ Lett.\  {\bf 60}, 1602 (1988); \\
  B.~S.~Balakrishna, A.~L.~Kagan and R.~N.~Mohapatra,
   ``Quark Mixings And Mass Hierarchy From Radiative Corrections'',
  Phys.\ Lett.\  B {\bf 205}, 345 (1988).

\bibitem{Balakrishna:1988xg}
  B.~S.~Balakrishna,
  ``Radiatively Induced Lepton Masses'',
  Phys.\ Lett.\  B {\bf 214}, 267 (1988).

\bibitem{Balakrishna:1988bn}
  B.~S.~Balakrishna and R.~N.~Mohapatra,
  ``Radiative fermion masses from new physics at TeV scale'',
  Phys.\ Lett.\  B {\bf 216}, 349 (1989).

\bibitem{Xing:2007fb}
  Z.~z.~Xing, H.~Zhang and S.~Zhou,
  ``Updated Values of Running Quark and Lepton Masses,''
  arXiv:0712.1419 [hep-ph].

\bibitem{Barr:1981wv}
  S.~M.~Barr,
  ``An SO(10) Model Of Fermion Masses,''
  Phys.\ Rev.\  D {\bf 24}, 1895 (1981); \\
  L.~Ferretti, S.~F.~King and A.~Romanino,
  ``Flavour from accidental symmetries,''
  JHEP {\bf 0611}, 078 (2006)
  [arXiv:hep-ph/0609047].

\bibitem{Babu:1990fr}
  K.~S.~Babu and R.~N.~Mohapatra,
  `` Permutation Symmetry And The Origin Of Fermion Mass Hierarchy'',
  Phys.\ Rev.\ Lett.\  {\bf 64}, 2747 (1990).

\bibitem{Miransky:1989ds}
 V.~A.~Miransky, M.~Tanabashi and K.~Yamawaki,
  ``Dynamical Electroweak Symmetry Breaking with Large Anomalous Dimension and
  t Quark Condensate,''
  Phys.\ Lett.\  B {\bf 221}, 177 (1989);
  ``Is the t Quark Responsible for the Mass of W and Z Bosons?,''
  Mod.\ Phys.\ Lett.\  A {\bf 4}, 1043 (1989).

\bibitem{Bardeen:1989ds}
  W.~A.~Bardeen, C.~T.~Hill and M.~Lindner,
  ``Minimal Dynamical Symmetry Breaking Of The Standard Model,''
  Phys.\ Rev.\  D {\bf 41}, 1647 (1990).

\bibitem{Dobrescu:1997nm}
  B.~A.~Dobrescu and C.~T.~Hill,
  ``Electroweak symmetry breaking via top condensation seesaw,''
  Phys.\ Rev.\ Lett.\  {\bf 81}, 2634 (1998)
  [arXiv:hep-ph/9712319]. \\ 
  R.~S.~Chivukula, B.~A.~Dobrescu, H.~Georgi and C.~T.~Hill,
  ``Top quark seesaw theory of electroweak symmetry breaking,''
  Phys.\ Rev.\  D {\bf 59}, 075003 (1999)
  [arXiv:hep-ph/9809470]. \\
  H.~J.~He, C.~T.~Hill and T.~M.~P.~Tait,
  ``Top quark seesaw, vacuum structure and electroweak precision
  constraints,''
  Phys.\ Rev.\  D {\bf 65}, 055006 (2002)
  [arXiv:hep-ph/0108041].


\bibitem{He:1989er}
  X.~G.~He, R.~R.~Volkas and D.~D.~Wu,
   ``Radiative Generation Of Quark And Lepton Mass Hierarchies From A Top Quark Mass Seed'',
  Phys.\ Rev.\  D {\bf 41}, 1630 (1990).

\bibitem{Babu:1990vx}
  K.~S.~Babu and R.~N.~Mohapatra,
  ``Top Quark Mass In A Dynamical Symmetry Breaking Scheme With Radiative B
  Quark And Tau Lepton Masses,''
  Phys.\ Rev.\ Lett.\  {\bf 66}, 556 (1991).

\bibitem{Rattazzi:1990wu}
  R.~Rattazzi,
  ``Radiative quark masses constrained by the gauge group only,''
  Z.\ Phys.\  C {\bf 52}, 575 (1991).

\bibitem{Wolfenstein:1983yz}
  L.~Wolfenstein,
  ``Parametrization Of The Kobayashi-Maskawa Matrix,''
  Phys.\ Rev.\ Lett.\  {\bf 51}, 1945 (1983).

\bibitem{Davidson:1993qk}
  S.~Davidson, D.~C.~Bailey and B.~A.~Campbell,
  ``Model independent constraints on leptoquarks from rare processes,''
  Z.\ Phys.\  C {\bf 61}, 613 (1994)
  [arXiv:hep-ph/9309310].

\bibitem{Cirigliano:2007xi}
  V.~Cirigliano and I.~Rosell,
  ``The Standard Model prediction for $R_{e/mu}^{(pi,K)}$,''
  arXiv:0707.3439 [hep-ph].

\bibitem{Kitano:2002mt}
  R.~Kitano, M.~Koike and Y.~Okada,
  ``Detailed calculation of lepton flavor violating muon electron  conversion
  rate for various nuclei,''
  Phys.\ Rev.\  D {\bf 66}, 096002 (2002)
  [Erratum-ibid.\  D {\bf 76}, 059902 (2007)]
  [arXiv:hep-ph/0203110].

\bibitem{Manohar:2006ga}
  A.~V.~Manohar and M.~B.~Wise,
  ``Flavor changing neutral currents, an extended scalar sector, and the  Higgs
  production rate at the LHC,''
  Phys.\ Rev.\  D {\bf 74}, 035009 (2006)
  [arXiv:hep-ph/0606172].

\bibitem{Gerbush:2007fe}
  M.~Gerbush, T.~J.~Khoo, D.~J.~Phalen, A.~Pierce and D.~Tucker-Smith,
  ``Color-octet scalars at the LHC,''
  arXiv:0710.3133 [hep-ph].
 
\bibitem{Dobrescu:2007yp}
  B.~A.~Dobrescu, K.~Kong and R.~Mahbubani,
  ``Massive color-octet bosons and pairs of resonances at hadron colliders,''
  arXiv:0709.2378 [hep-ph].
  
\bibitem{Gresham:2007ri}
  M.~I.~Gresham and M.~B.~Wise,
  ``Color Octet Scalar Production at the LHC,''
  Phys.\ Rev.\  D {\bf 76}, 075003 (2007)
  [arXiv:0706.0909 [hep-ph]].

\bibitem{Babu:1989tv}
  K.~S.~Babu, B.~S.~Balakrishna and R.~N.~Mohapatra,
  ``Supersymmetric model for fermion mass hierarchy'',
  Phys.\ Lett.\  B {\bf 237}, 221 (1990); \\
  S.~Nandi and Z.~Tavartkiladze,
  ``A New Extensions of MSSM: FMSSM,''
  arXiv:0804.1996 [hep-ph].

\end{thebibliography}
\end{document}